\documentclass{pas}

\usepackage{aas_macros}
\usepackage{multirow}
\usepackage{physics}
\usepackage{chemformula}
\usepackage{hyperref}

\begin{document}

\lefttitle{Evolutionary Pathways of Double White Dwarf Binaries}
\righttitle{S. Roy \& S. Kalita}

\jnlPage{1}{10}
\jnlDoiYr{2026}
\doival{10.1017/pasa.xxxx.xx}

\articletitt{Research Paper}

\title{Population synthesis of double white dwarfs: evolutionary effects on system properties}

\author{\gn{Sreeta} \sn{Roy}$^{1}$ and \gn{Surajit} \sn{Kalita}$^{1}$}

\affil{$^1$Astronomical Observatory, University of Warsaw, Al. Ujazdowskie 4, PL-00478 Warszawa, Poland}

\corresp{S. Roy, Email: sroy@astrouw.edu.pl}

\citeauth{S. Roy and S. Kalita, Population synthesis of double white dwarfs: evolutionary effects on system properties. {\it Publications of the Astronomical Society of Australia} {\bf 00}, 1--10. https://doi.org/10.1017/pasa.xxxx.xx}

\history{(Received xx xx xxxx; revised xx xx xxxx; accepted xx xx xxxx)}

\begin{abstract}
Double white dwarf (DWD) binaries are natural outcomes of binary stellar evolution and key sources for future space-based gravitational wave (GW) observatories such as Laser Interferometer Space Antenna (\textit{LISA}). We investigate how different binary interaction channels shape the physical and orbital properties of DWD systems, focusing on component masses, orbital separations, core compositions, and mass transfer rates. Using the binary population synthesis code \textsc{compas}, we evolve $10^7$ binaries with physically motivated initial distributions of binary parameters. Our simulations reproduce the strong bimodality in the final orbital separations, including a pronounced deficit of systems around $100-500 \rm\,R_\odot$, arising from distinct evolutionary pathways: wide DWDs predominantly originate from stable Roche lobe overflow (RLOF), while close DWDs form through unstable RLOF leading to at least one common envelope (CE) phase. Moreover, we show that the core compositions of WDs provide a powerful tracer of evolutionary history: He-core WDs are strongly concentrated in close systems, whereas CO-core WDs span the full separation range and exhibit a small mass gap in wide binaries. We further identify a correlation between the donor mass transfer rate and the final orbital separation, highlighting the impact of non-conservative mass transfer on the resulting orbital configuration of DWD systems. These results underscore the links among evolutionary channels, chemical composition, and mass transfer rates; thereby provide a unique framework for interpreting \textit{Gaia} DWD samples and forecasting the joint electromagnetic and GW population accessible to \textit{LISA}.
\end{abstract}

%\begin{keywords}
%white dwarfs – binaries: close – binaries: general – stars: evolution – gravitational waves – methods: numerical
%\end{keywords}

\maketitle

\section{Introduction}

White dwarfs (WDs) are the most abundant stellar remnants constituting a significant component of the stellar population in the Universe. This makes them crucial probes for studying the structure and evolution of compact objects~\citep{2001PASP..113..409F,2010A&ARv..18..471A}. A considerable fraction of WDs exist in binary systems (a system comprising of two WDs), commonly referred to as double WDs (DWDs), which play a pivotal role in several astrophysical contexts. These systems often undergo multiple distinct mass transfer phases, making them ideal laboratories for investigating binary stellar evolutions. Furthermore, the coalescence of DWDs within such systems is one of the leading progenitor models for Type Ia supernovae~\citep{1984ApJ...277..355W,2010ApJ...710.1310M,2015ApJ...807..105S,2018MNRAS.480.4519C}. Additionally, DWDs are expected to dominate the low-frequency gravitational wave (GW) spectrum, potentially generating a stochastic background that can obstruct other GW source signals~\citep{1987ApJ...323..129E,2001A&A...375..890N,2019MNRAS.490.5888L}. Moreover, Galactic DWD binaries create a dominant GW foreground for Laser Interferometer Space Antenna (\textit{LISA}) with mass-transferring systems significantly enhancing the noise at higher frequencies while still leaving thousands of resolvable sources~\citep{2010ApJ...717.1006R}. Finally, analysing close DWDs, especially considering the binary evolution theory, provides a way to test and refine cooling models for low-mass WDs~\citep{2000MNRAS.316...84S,2016A&A...595A..35I}.

The formation of DWDs has been studied extensively using both analytical~\citep{1986ApJ...311..753I,1987ApJ...313..727I} and numerical population synthesis methods~\citep{1988Ap&SS.145....1L,1993ASPC...38..211Y,1995MNRAS.272..800H,1997ApJ...475..291I,1998MNRAS.296.1019H}. These studies primarily differ in their assumptions about key evolutionary processes, such as common envelope (CE) ejection, mass transfer stability, and angular momentum loss, which lead to variations in predicted population characteristics~\citep{2022MNRAS.511.5936K,2023MNRAS.518.3966S,2024A&A...692A.165T,2025A&A...699A.172V}. A cross-comparison of these models provides valuable insights into how such modelling choices affect the predicted properties of DWD populations.

In recent years, the catalogue of known DWDs has grown substantially owing to surveys such as the Extremely Low Mass survey and Double-Lined DWD survey, which have collectively identified DWD systems with well-constrained masses and orbital periods~\citep{2020ApJ...889...49B,2024MNRAS.532.2534M,2025MNRAS.541.3494M}. This expanded dataset allows for a more precise comparison between theoretical predictions and observed populations. In particular, the discovery of ultracompact systems, such as J0526+5934 with an orbital period of just 20.5\,mins~\citep{2024A&A...686A.221R}, has enabled stringent tests of binary evolution at the shortest orbital separations.

In parallel, \textit{Gaia}-based analyses have improved estimates of the local DWD fraction, revealing that approximately 1.2\% of resolved WDs within 100\,pc reside in DWD systems, while approximately 6.3\% are in resolved binaries with the companion being a main-sequence star~\citep{2022MNRAS.511.5462T}. Furthermore, forward modelling efforts for \textit{LISA} GW mission predict that up to approximately 16\,000 Galactic DWDs may be individually detectable over a 10\,yr observing run~\citep{2022MNRAS.511.5936K}, offering a unique opportunity to study the compact binary population through GW observations.

In this study, we utilise the binary population synthesis code \textsc{compas} (Compact Object Mergers: Population Astrophysics and Statistics)\footnote{\url{https://compas.science/}} to investigate the influence of different binary evolutionary scenarios on the formation properties and accretion histories of DWDs. By systematically exploring key physical processes, such as Roche lobe overflow (RLOF), mass transfer stability, and core composition of WDs, we assess their impacts on the resulting WD core types, orbital properties, and mass distributions. In particular, we recover the previously reported bimodal separation distribution \citep{2022MNRAS.511.5936K, 2022MNRAS.515.1228K} with intermediate under-density in DWD orbital separations, and we extend earlier work by (i) providing a unified analysis that links formation channel (RLOF-only vs. CE-involving), WD core type (He/CO/ONe), and mass transfer rate within a single \textsc{compas}-based framework, and (ii) quantifying how the separation gap responds to changes in the CE structural parameter. Our results provide a physically interpretable map between the observed properties of DWDs and their underlying interaction histories, and thus offer a concrete framework for interpreting future detections, particularly in the context of upcoming space-based GW observatories like \textit{LISA}.

The remainder of this paper is organised as follows. In Section~\ref{Sec2}, we review the key aspects of binary stellar evolution and the formation of DWDs. Section~\ref{Sec3} provides our key results, and Section~\ref{Sec4} discusses the results and compares them with the previous literature. Finally, concluding remarks are provided in Section~\ref{Sec5}.

%--------------------------------------------------------------------
\section{Revisiting binary evolution physics}\label{Sec2}

The evolutionary pathways of binary stellar systems are similar to those of single stars when their orbital separations are sufficiently wide that neither star fills its Roche lobe during their evolution. In this case, the stars evolve independently, without undergoing mass transfer or tidal interactions. Therefore, in wider binaries, each star evolves independently according to the single-star evolution process. However, in close binaries (systems with separations small enough that at least one star is expected to fill its Roche lobe within the Hubble time), gravitational interactions make their evolutionary paths interdependent, leading to complex phenomena that significantly reshape the evolution of both stars and the system itself.

A fundamental process in close binary stellar evolution is mass transfer, which is initiated when one star expands to fill its Roche lobe, a region gravitationally bound to the star. The Roche lobe radius for the donor star ($R_L$) can be approximated by the following relation given by \cite{1983ApJ...268..368E} as
\begin{equation}
    \frac{R_L}{a} = \frac{0.49q^{2/3}}{0.6q^{2/3} + \ln({1 + q^{1/3}})},
\end{equation}
where $a$ is the semi-major axis of the binary and $q$ is the mass ratio of donor to accretor.

The stability of mass transfer depends critically on the structure of the donor star and mass ratio. Stable mass transfer occurs if the donor star and its companion have comparable masses or the donor has a radiative envelope. In this case, matter is transferred through the inner Lagrangian point and accreted by the companion over a long timescale. This scenario allows both stars to adjust with the changing mass ratio without significant disruption. On the other hand, unstable mass transfer occurs if the donor is a convective giant or significantly more massive than its companion. In this case, the companion cannot efficiently accrete the incoming material, leading to envelope engulfment~\citep{1976IAUS...73...75P} and CE phase. However, the stability of mass transfer remains one of the open problems in binary evolution. The location of the stability boundary is governed by multiple physical factors, such as the donor’s internal structure (whether it has a radiative or convective envelope, or is partially degenerate), the binary mass ratio, the timescale of mass transfer (nuclear, thermal, or dynamical), and a number of poorly constrained processes such as angular momentum loss via outflows, tides, or magnetic braking. Recent studies have shown that even moderate changes in these assumptions can substantially alter the predicted stability threshold, particularly for giants and partially stripped progenitors. Consequently, the standard guideline of that donors with radiative envelopes undergo stable mass transfer while those with convective envelopes experience unstable mass transfer is now recognized as an oversimplification~\citep{2024arXiv241117333G}. This uncertainty in the stability boundary implies that the relative rates of stable RLOF and CE evolution in any rapid population synthesis model should be interpreted as model-dependent rather than as precise predictions.

In this work, we adopt the default \textsc{compas} stability criteria without recalibration, which are based on the rapid binary evolution models by \cite{2002MNRAS.329..897H}. Our population-level predictions are therefore rely on these prescriptions. We partially address this model dependence by studying the sensitivity of our results to the changes in significant parameters such as the degree of non-conservative mass transfer and the angular momentum lost through outflows. These variations can alter the occurrence of stable RLOF versus CE evolution. A comprehensive treatment, which would involve the combination of complete stellar-structure models with the mass transfer process on thermal and dynamical timescales, is not within the realm of a rapid population synthesis study. Nevertheless, we discuss these outcomes which remain robust across a plausible range of input parameters. Therefore, our research should be interpreted as an exploratory study of how different interaction channels imprint themselves on DWD population, rather than as a fully calibrated model of the observed systems.

The change in orbital separation during the CE phase is typically modelled using the energy formalism~\citep{1984ApJ...277..355W}, given by
\begin{equation}
    \alpha_\text{CE}\left(\frac{GM_\text{c,1}M_{2}}{2a_\text{f}} - \frac{GM_1M_2}{2a_\text{i}}\right) = \frac{GM_1M_\text{env,1}}{\lambda_\text{CE} R_1},
\end{equation}
where $\alpha_\text{CE}$ is the CE efficiency, $M_\text{c,1}$ is the core mass of the donor, $M_\text{env,1}$ is the envelope mass, $a_\text{i}$ and $a_\text{f}$ are the initial and final semi-major axis, $R_1$ is the radius of the donor, $M_1$ and $M_2$ are the mass of the donor and the companion, and $\lambda_\text{CE}$ is the parameter that describes the structure and binding energy of the donor star’s envelope. While the energy formalism is commonly applied in rapid population synthesis, alternate models exist. The angular momentum based prescription proposed by \cite{2000A&A...360.1011N} links orbital contraction to angular momentum loss rather than energy considerations. These two formalisms represent competing physical interpretations of the CE process, and the choice between them can substantially alter the predicted post-CE orbital separation distribution. In the present work, we adopt only the energy formalism and do not implement the angular-momentum prescription, so our post-CE separation distributions should be interpreted within this framework.
During this brief and dynamic CE phase, drag forces within the envelope cause the stellar cores to spiral inward, which may lead to ejection of the envelope and formation of a close binary. If the envelope ejection is inefficient, it may lead to a stellar merger. Thus, the outcome of this CE phase determines whether the system evolves into a compact binary or coalesces.

These complex evolutionary pathways lead to the formation of diverse compact binary systems and unusual stellar phenomena, including cataclysmic variables, X-ray binaries, Algol-type systems, and blue stragglers, which often cannot be explained by a single stellar evolution alone~\citep{1985ibs..book.....P,1996ASIC..477.....W,2009ApJ...697.1048P}. Although many details of close binary evolution are uncertain, a robust theoretical framework now exists that enables us to track how binaries progress through different evolutionary phases. In the context of DWD formation, progenitor properties, such as stellar mass, core composition, orbital period or semi-major axis, and eccentricity, determine the sequence of interactions that may eventually produce a binary system containing two WDs. Mapping of these evolutionary pathways is essential for interpreting the observed DWD population and predicting the outcomes of future GW surveys.
%--------------------------------------------------------------------
\subsection{Double white dwarf formation}\label{sec2.1}

Close DWDs, which in our simulations correspond to orbital separations of $\sim 0.05 -100 \rm\,R_\odot$ (i.e. periods from minutes to days/months), are predominantly formed through two widely discussed families of evolutionary channel. In the first case, the primary star evolves to fill its Roche lobe, initiating stable mass transfer to the secondary star, which may or may not accrete the material. This phase produces an intermediate binary consisting of the first-formed WD and a secondary star (main-sequence or asymptotic giant branch). Subsequently, the secondary star fills its Roche lobe, triggering a second mass transfer episode. If this second mass transfer phase remains dynamically stable, the initially more massive primary has already become a WD, so mass is now transferred from the more massive secondary to the lighter WD (a mass ratio greater than unity). This configuration generally tends to shrink the orbit, and in our non-conservative models, additional angular momentum is carried away by the expelled material. As a result, binaries that undergo two stable RLOF episodes (RLOF+RLOF) can also produce close DWDs with separations of only a few solar radii, even without ever entering a CE phase. If the donor is a convective giant or significantly more massive than the WD companion, this phase becomes dynamically unstable~\citep{1976IAUS...73...75P,1979IAUS...83..401T,1984ApJ...277..355W,1993PASP..105.1373I}. During this phase, the core of the donor and its companion spiral together within the shared envelope because of drag forces, which release orbital energy. This energy heats and ultimately expels the envelope, resulting in a significant reduction in the orbital separation. This explains how DWDs with separations of only a few solar radii form from initially wider binaries~\citep{2012ApJ...744...12W,2013A&ARv..21...59I}. 

In the other scenario, both stars undergo consecutive CE phases, each initiated by an unstable RLOF. In our rapid binary-evolution framework, every CE event is triggered by dynamically unstable mass transfer (or an equivalent envelope-engulfment instability), so episodes of unstable RLOF and CE phases are effectively interchangeable in what follows. In this case, changes in the orbital separation during these phases are typically described by the balance between the orbital energy lost and the binding energy required to eject the envelope~\citep{1976IAUS...73...75P,1984ApJ...277..355W,1993PASP..105.1373I,2021A&A...648L...6P}. This pathway typically produces tighter DWD systems as both episodes of envelope ejection efficiently shrink the orbit. These mechanisms account for the observed populations of close DWDs and their diverse orbital properties~\citep{2016ApJ...818..155B,2026NewA..12202477N}. The stability of the mass transfer, efficiency of envelope ejection, and initial binary parameters critically determine the final configuration of the system~\citep{2021A&A...651A.100O}. In summary, the DWDs in our simulations arise from a small number of recurring interaction histories, most commonly either two phases of stable RLOF or sequences that include at least one episode of unstable mass transfer leading to a CE (e.g. RLOF+CE, CE+RLOF, or CE+CE). For the purposes of the population analysis below, we therefore group systems into two broad families: a purely stable RLOF channel, comprising binaries that never enter a CE (zero CE episodes), and a CE involving channel, comprising all binaries that experience at least one CE episode, independent of the exact order of stable and unstable mass transfer. Wide DWDs in our models are predominantly produced by the RLOF only channel, whereas the tightest systems almost always originate from CE involving histories. 

Another possible pathway for the formation of extremely low-mass (ELM) DWDs involves a two-stage process: the system first undergoes a CE phase triggered by unstable mass transfer, leading to the formation of a compact WD + MS binary. This is then followed by a stable RLOF phase, during which the MS companion transfers mass to the WD, producing a ELM WD \citep{2019ApJ...871..148L}. In our simulations, any DWDs formed in this way would belong to the CE involving family, but we do not attempt to isolate or quantify this specific channel separately, any systems formed in this way would simply be included within the broader CE involving population in our analysis.

%--------------------------------------------------------------------
\subsection{Binary population synthesis with \textsc{compas} simulation}

To generate synthetic binary populations of WDs, we use \textsc{compas} v02.50.00~\citep{2022ApJS..258...34R,2025ApJS..280...43T}, an open-source rapid binary population synthesis (BPS) code. \textsc{compas} simulates the evolution and interaction of millions of stellar binaries using a grid of parametrised physical prescriptions. Similar to other BPS codes such as \textsc{bse}~\citep{2002MNRAS.329..897H}, \textsc{StarTrack}~\citep{2008ApJS..174..223B}, \textsc{Binary C}~\citep{2004MNRAS.350..407I}, \textsc{SeBa}~\citep{2012A&A...546A..70T}, and \textsc{cosmic}~\citep{2020ApJ...898...71B}, \textsc{compas} is computationally efficient, enabling rapid exploration of large parameter spaces and statistical studies of binary populations. This efficiency is achieved through simplified evolutionary prescriptions, enabling the evolution of millions of binaries in just a few CPU hours, although at the cost of not resolving the detailed internal structure of individual stars captured by 1D stellar evolution codes such as \textsc{mesa} or other hydrodynamic simulations~\citep{2018PASA...35...31P}. While \textsc{compas} is valuable in generating binary populations, it does not resolve detailed stellar structure, hydrodynamics, or dynamical interactions in dense environments, and relativistic effects are included only through approximate orbital decay prescriptions. As a result, predicted binary properties and merger rates are sensitive to model assumptions and should be interpreted statistically rather than as precise predictions for individual systems~\citep{2023MNRAS.524..245R,2024MNRAS.534.3506R}.

While \textsc{compas} was primarily applied to study the formation and evolution of neutron stars and black holes~\citep{2022MNRAS.516.5737B,2022MNRAS.517.4034S,2022ApJ...937..118W}, we extend its application to low-mass binaries in this study, focusing specifically on the formation channels and population properties of DWD systems. Our study thus explores the evolution of binaries that produce WDs and establishes a framework for future, more detailed investigations of low and intermediate mass stellar evolution and for calibration against observed DWD populations. Moreover, in the present version of \textsc{compas}, magnetic braking for low-mass stars is not incorporated which mostly acts while low mass stars with convective envelopes are still on the main sequence. This would have made close orbit binaries tighten faster and thereby affecting the formation efficiency and orbital period distribution for low-mass systems~\citep{2024A&A...682A..33B}; whereas it has little impact on the wide DWDs.

In this study, we use \textsc{compas} to generate a large population of DWDs by sampling initial binary parameters from observationally motivated distributions. The primary masses are drawn from a Salpeter initial mass function in the range $0.5-15 \rm\,M_{\odot}$, and the secondary masses are set by a flat initial mass ratio distribution~\citep{1955ApJ...121..161S}. This mass range includes primaries that are massive enough to form WDs within a Hubble time while avoiding very low-mass stars that would not evolve off the main sequence. Systems in which either component undergoes core collapse are excluded from our DWD sample. The semi-major axes are sampled from FLATINLOG (i.e. flat in $\log a$) distribution over $0.01-10\rm\, AU$ covering both very close binaries that are almost guaranteed to interact via RLOF and moderately wide systems that can still undergo interaction once one or both stars evolve to giant radii. Much wider binaries, which are unlikely to experience strong interaction or to form close DWDs, are not considered here. Eccentricities are drawn from a thermal distribution of $f(e)=2e$ with $e$ being the eccentricity~\citep{1975MNRAS.173..729H}. All binaries are assumed to have a fixed metallicity ($Z=0.014$) throughout the evolution from progenitor binaries to the formation of DWDs. For our main simulation set, we evolve $10^7$ binaries using the default rapid-synthesis mode of \textsc{compas}, identifying DWDs as systems in which both stars become WDs with He, CO, or ONe-cores~\citep{2002MNRAS.329..897H}. In \textsc{compas}, WDs are classified as He, CO, or ONe-core objects according to their prior core-burning history (e.g. whether He has been exhausted and whether carbon ignition has occurred), rather than a strict cut on the final WD mass. This is important when interpreting low-mass CO tagged objects and massive CO/ONe remnants in our population. In our simulations, each binary is evolved from zero-age main sequence (ZAMS) up to a maximum age equal to the Hubble time (the default maximum evolution time in \textsc{compas}, $\sim$13.8\,Gyr), or until both components have become compact remnants. This enables us to analyse their orbital separations, masses, compositions, and mass transfer rates to connect formation channels with observable properties.

Unless otherwise specified, we adopt standard \textsc{compas} physical prescriptions with CE efficiency parameters $\alpha_\text{CE}=1.0$ and $\lambda_\text{CE} = 0.1$. We do not attempt to recalibrate these parameters to match the observed DWD population. Instead, we treat them as plausible but uncertain choices and also explore how varying $\lambda_\text{CE}$ affects the separation distribution in Section~\ref{Sec3}. The resulting synthetic population forms the basis of our analysis in subsequent sections.

%--------------------------------------------------------------------
\section{Results}\label{Sec3}
\subsection{Effects of evolutionary channels on DWD properties}

\begin{figure}[htbp] 
    \centering
    \subfigure[~Primary mass.]{\includegraphics[scale=0.5]{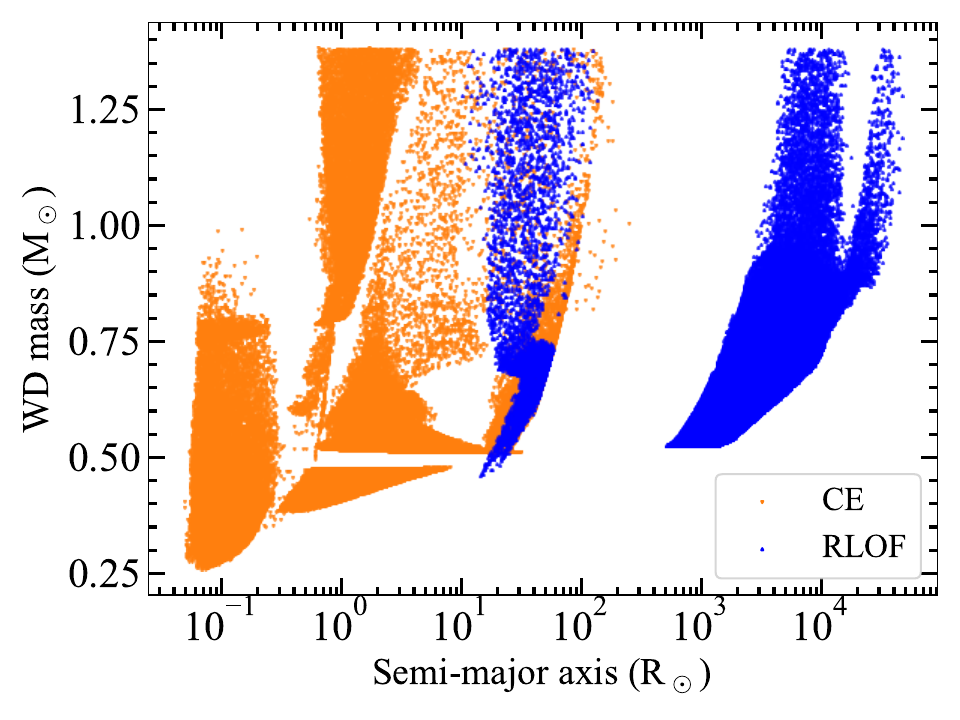}}    
    \subfigure[~Secondary mass.]{\includegraphics[scale=0.5]{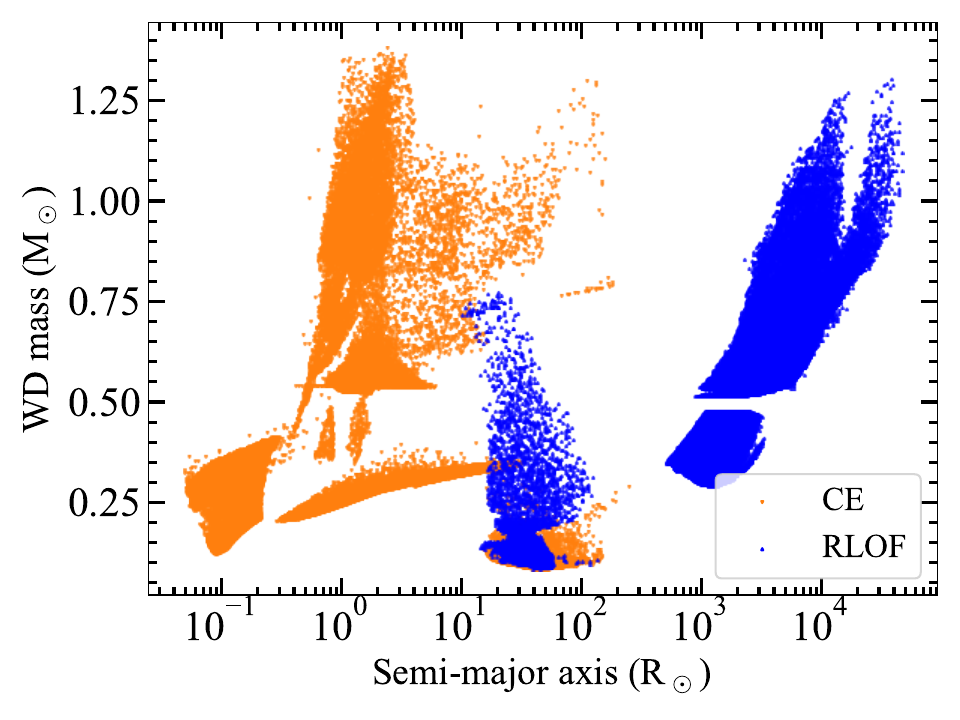}}    
    \caption{Final WD mass as a function of orbital separation for DWDs for $\alpha_\text{CE}=1.0$ and $\lambda_\text{CE} = 0.1$. Systems evolving solely through stable RLOF are shown in blue, while those involving at least one CE phase are shown in orange. Panel~(a) represents the more massive WD and panel~(b) represents the less massive companion in each binary system.}
    \label{Fig: CE vs. RLOF}
\end{figure}

Our simulations confirm that the evolutionary pathways leading to DWDs can usefully be grouped into two broad categories: (i) systems that evolve solely through stable RLOF and never enter a CE phase, and (ii) systems that experience at least one CE episode. These two interaction channels together account for the large majority of DWDs in our models. In addition to these interaction-driven pathways, our models also produce a small number of DWDs that never undergo sustained RLOF or CE phase. These wide and effectively non-interacting systems are not included in the two formation channels defined above and are not considered hereinafter. Each channel has distinct outcomes on DWD properties, especially in their masses and orbital separation. Figure~\ref{Fig: CE vs. RLOF} illustrates this by plotting the final semi-major axis against the individual WD mass formed from each channel. Each point in the plot represents a WD in the binary system, with panel~(a) showing the more massive (primary) WD and panel~(b) the less massive (secondary) WD in each binary.

Both panels of Figure~\ref{Fig: CE vs. RLOF} reveal a bimodal distribution in the orbital separation of DWDs, reflecting their evolutionary histories.
\begin{itemize}
    \item Systems evolving solely through stable RLOF mass transfer tend to remain at wide orbital separations, with final semi-major axes predominantly exceeding $500\rm\,R_\odot$, and in some cases extending beyond $10^4\rm\,R_\odot$, along with a relatively small fraction of RLOF only systems at smaller separations. These close RLOF only binaries arise from net orbital shrinkage during the second stable RLOF episode (see aforementioned discussion of Section~\ref{sec2.1}). The component WD masses show a broad distribution, spanning approximately $0.5-1.4\rm\,M_\odot$ for primary and $0.1-1.15\rm\,M_\odot$ for secondary component. This reflects the diversity of initial binary parameters and mass transfer histories. A sharp lower mass cut-off for the primary component is evident at approximately $0.5\rm\,M_\odot$, whereas a small gap is present for the secondary component around that value. In the next section, we show that these features arise from the distribution of chemical compositions of WDs.
    \item In contrast, systems undergoing at least one CE phase end up in much closer orbits, with their final semi-major axes mainly below $100\rm\,R_\odot$. Their orbital separations are orders of magnitude smaller, while the WD masses are more spread out compared to the stable RLOF only channel. No significant mass gap is observed in this population. These CE involving systems arise from the various RLOF+CE, CE+RLOF, and CE+CE sequences discussed in Section~\ref{Sec2}, but regardless of the detailed order of events, the presence of at least one CE episode is what drives the strong orbital contraction.

\end{itemize}

The key feature of both panels in Figure~\ref{Fig: CE vs. RLOF} is the well-defined gap in orbital separation between approximately $100$ and $500\rm\,R_\odot$. The lack of systems at this intermediate separation in our models highlights a characteristic imprint of binary interaction: the mechanisms that dominate DWD formation preferentially yield either wide or close binaries, with a few systems in between. Stable mass transfer is typically conservative or weakly non-conservative \citep{1997A&A...327..620S}, often resulting in orbital expansion, whereas the CE phase is highly efficient at extracting orbital energy, inducing a significant contraction of the orbit. The consistency of this bimodal separation distribution across both primary and secondary WD populations in our simulations strongly suggests that the type of binary interaction (purely stable RLOF versus at least one CE episode) is the dominant factor in setting the final orbital separation of DWD systems. We show later in this section that varying the CE binding parameter $\lambda_\text{CE}$ modifies the depth and width of the gap but not its overall presence.

Current DWD catalogues are majorly incomplete due to strong observational selection effects, particularly for close binaries. However, there is emerging evidence of DWDs at intermediate separations. In particular, \cite{2022MNRAS.511.5936K,2022MNRAS.515.1228K} combined population synthesis models with \textit{Gaia} selected wide DWDs and showed that their best-fitting models require relatively fewer systems at separations near 1\,AU in order to match the observed orbital separation distribution. This characteristic scale is consistent with our inference of gap at $100-500\rm\,R_\odot$($\approx 0.46-2.32$\,AU). A detailed, one-to-one comparison with the observed DWD population, including a full analysis of observational selection effects, is beyond the scope of this study. We therefore restrict our discussion to emphasizing the order-of-magnitude agreement between our predicted gap and the under-density inferred from \textit{Gaia}-calibrated Galactic DWD population models relevant for \textit{LISA}.

The physical origin of the gap is associated with binary interactions during post main-sequence evolution, where the maximum radius ($R_\text{max}$) of WD progenitors during RGB/AGB evolution can reach upto a few tens to a few hundred solar radii and it eventually sets the characteristic separation scale $a_\text{crit} \simeq R_\text{max}/f(q)$ with $f(q)\equiv R_L/a$. Systems with $a\lesssim a_\text{crit}$ inevitably undergo RLOF~\citep{1988ApJ...334..688L}. If mass transfer is stable and only weakly non-conservative, the orbit widens, yielding final separations $\gtrsim500\rm\,R_\odot$. A small subset of RLOF-only systems can experience two stable mass transfer phases. During the second episode, mass transfer from the initially less massive star onto the first-formed WD leads to orbital shrinkage, producing relatively close DWDs with separations of a few tens to $\sim 100 \rm\,R_\odot$ without a CE phase. In contrast, systems that undergo at least one CE phase experience dynamically unstable mass transfer, resulting in substantial orbital energy loss and final separations $\lesssim100\rm\,R_\odot$.

%--------------------------------------------------------------------
\subsection{Effects of chemical compositions on DWD properties}
\begin{figure}[htpb]
    \centering
    \subfigure[~Primary mass.]{\includegraphics[scale=0.5]{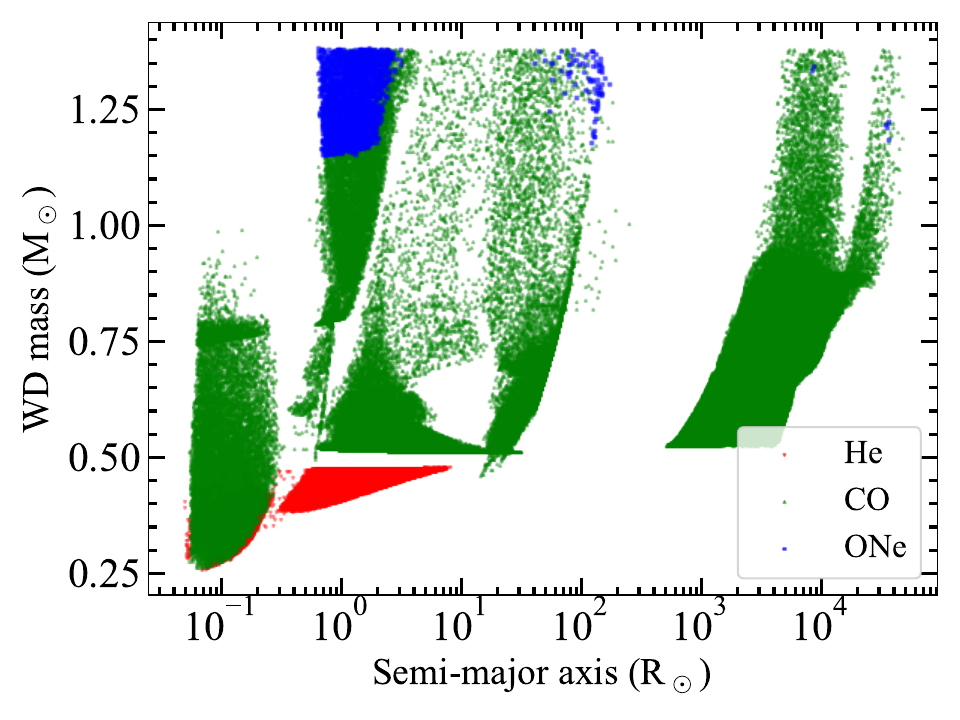}}    
    \subfigure[~Secondary mass.]{\includegraphics[scale=0.5]{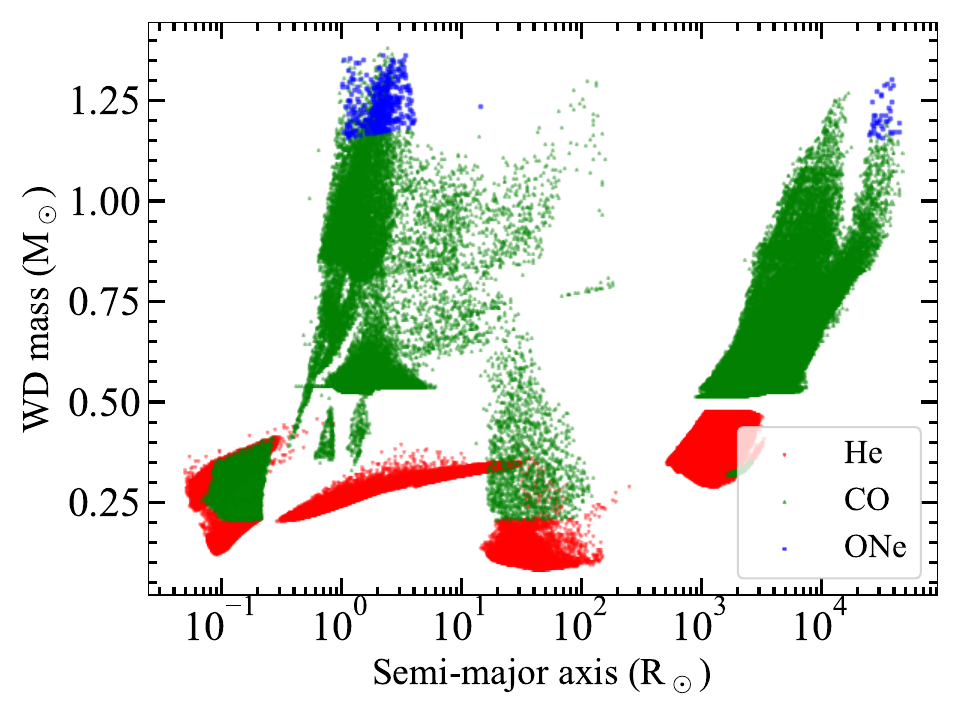}}    
    \caption{Same as Figure~\ref{Fig: CE vs. RLOF} except each WD is marked according to their core composition.}
    \label{Fig: ST}
\end{figure}

To explain the features observed in the orbital separation distributions, especially in tight binaries, we examine the relationship between final orbital parameters and WD core composition. Figure~\ref{Fig: ST} illustrates the final semi-major axis and mass for DWDs categorised by their core composition (He, CO, or ONe). Similar to that in the previous case, panel~(a) corresponds to the more massive WD in each binary, while panel~(b) shows the less massive component.

Our analysis reveals a strong correlation between core type, mass, and orbital separation.
\begin{itemize}
    \item He-core WDs are confined to low mass ends ($\lesssim0.5\rm\, M_\odot$), exhibiting a upper mass cut-off. This upper mass limit is consistent with the \textsc{compas} remnant-typing scheme described above, where WDs are classified according to whether their stripped cores ignite He and begin building a CO core. The heavier components in a DWD system are found exclusively in close binaries, originating from a CE evolution. In contrast, secondary He-core WDs appear in both wide and close binaries, indicating diverse formation pathways for the lower-mass component. 
    \item CO-core WDs dominate the population and span across intermediate to higher mass ($\gtrsim 0.5\rm\,M_\odot$) with a relatively small fraction at lower mass. This is because of \textsc{compas} classification of WDs: once a stripped core ignites He and starts building a CO-core, the remnant is tagged as a CO-core WD even if He burning is incomplete and the total mass remains relatively low. As a result, the region around $0.5\rm\,M_\odot$ marks the transition from He-core to CO-core WDs in the models. CO-core WDs are found across the entire spectrum of orbital separations. In wide binaries, the more massive WD is invariably a CO-core WD, as its progenitor must have evolved to at least the tip of the red giant branch (RGB) to ignite He. This behaviour is consistent with standard expectations for He-CO transitions in low and intermediate mass stars, and with previous population synthesis results
    \item ONe-core WDs are very few in numbers and occupy the highest mass ($\gtrsim1.2\rm\, M_\odot$). They are mostly found at intermediate separations and originated from high mass progenitors. In a few systems, the CO-core companion can be comparable in mass or even slightly more massive than the ONe WD, reflecting earlier accretion driven growth of the CO remnant and the fact that the CO-ONe boundary in \textsc{compas} is set by core ignition history rather than a sharp mass threshold.
\end{itemize}

Figure~\ref{Fig: ST}(b) shows the small mass gap for a stable RLOF channel in the secondary WD population (Figure~\ref{Fig: CE vs. RLOF}b) is associated with the core composition, where He-core WDs exhibit lower masses and CO-core WDs exhibit higher masses. In our models, we do not find any double He-core WD systems in wide binaries formed exclusively through the stable RLOF channel. The formation of such systems requires both stars to undergo envelope stripping during the RGB phase via stable mass transfer, with the evolution precisely stopped prior to He ignition. This process requires a highly specific range of initial stellar masses, orbital separations, and mass transfer efficiencies. In practice, the secondary star typically evolves to a more advanced evolutionary stage, resulting in a CO-core WD, thereby preventing the formation of double He-core wide binaries. The absence of such systems in our population should therefore be interpreted as a consequence of our adopted stability criteria and mass transfer prescriptions, rather than as a strict prediction that wide double He-core DWDs cannot form in nature.

%--------------------------------------------------------------------
\subsection{Effects of mass transfer rates on DWD properties}
\begin{figure}[htbp] 
    \centering
    \subfigure[~Primary mass.]{\includegraphics[scale=0.5]{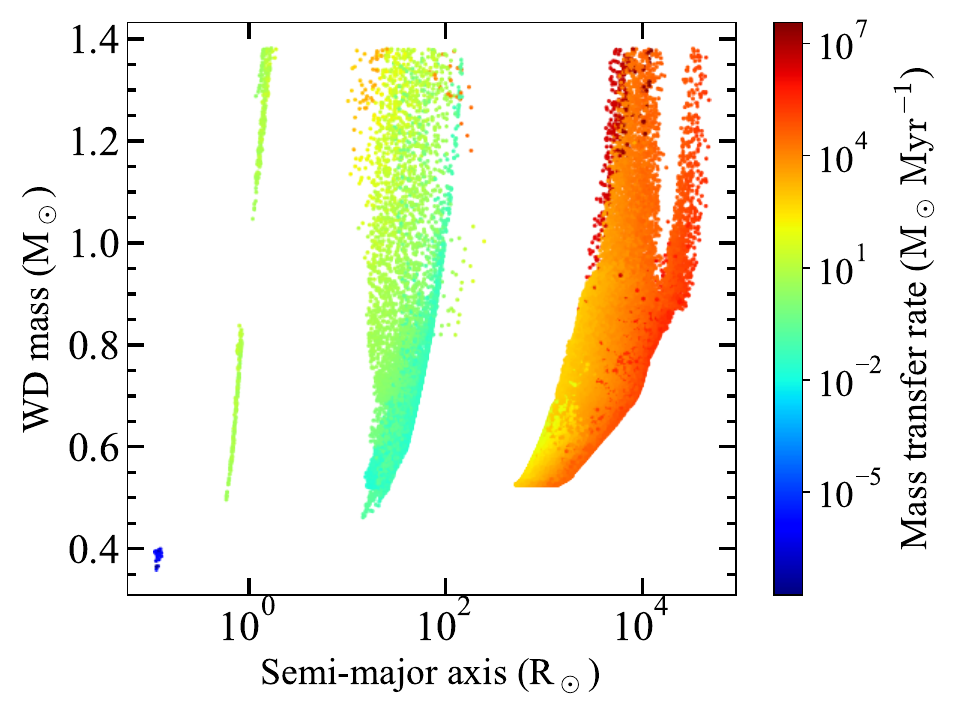}}    
    \subfigure[~Secondary mass.]{\includegraphics[scale=0.5]{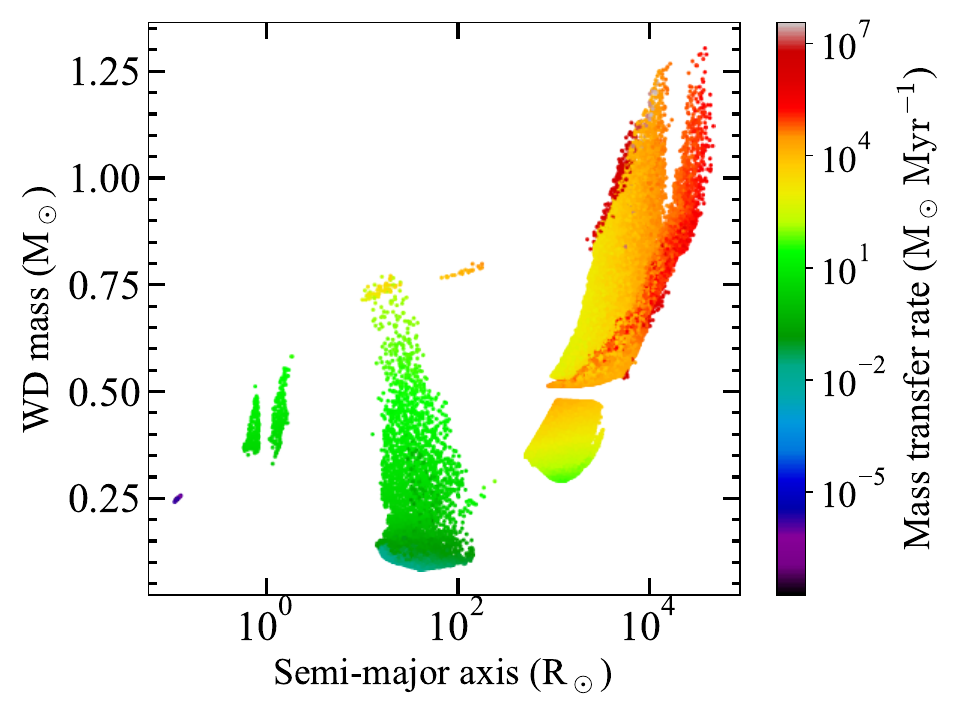}}    
    \caption{Same as Figure~\ref{Fig: CE vs. RLOF} except that donor mass transfer rates during the last stable mass transfer episode is shown in the colour bar.}
    \label{Fig: Accretion}
\end{figure}

To further understand the features observed in the orbital separation distributions, especially in tight binaries, we investigate the role of mass transfer efficiency. In particular, we examine how the efficiency of stable mass transfer and the associated angular-momentum loss map onto the final orbital separations, which is expected to produce a correlation between mass transfer rate and post-interaction separation in standard binary-evolution theory. Figure~\ref{Fig: Accretion} shows the final semi-major axis and mass for DWDs, with the colour bar indicating the mass transfer rate during the most recent RLOF episode. Note that for this analysis, we restrict the sample to systems that underwent at least one stable RLOF mass transfer phase. This removes many WDs including the primary components with masses $\lesssim0.5\rm\,M_\odot$, which are typically formed without any stable RLOF episode. 

The key observations from the plots are as follows.
\begin{itemize}
    \item Wide binaries are characterized by the highest mass transfer rates. This is a direct consequence of the \textsc{compas} prescription for stable mass transfer. When a star overflows its Roche lobe, \textsc{compas} initiates stable mass transfer on a thermal (or, for some systems, nuclear) timescale. In these wide systems, the mass transfer is often nearly conservative, brief, and intense, resulting in high donor mass transfer rates. This interaction does not significantly alter the orbit, allowing both stars to subsequently evolve independently and form a wide DWD.
    \item Close and interacting binaries, on the other hand, can evolve into DWDs by mass transfers with lower to higher mass transfer rates. Comparing Figure~\ref{Fig: Accretion} with Figure~\ref{Fig: ST}, it is evident that for close DWDs with the primary component being a CO-core WD and the secondary being a He-core WD, the mass transfer rate is the lowest. For double CO-core WD in close binaries, the mass transfer rate is relatively higher, and it increases as the final orbital separation decreases. This anti-correlation between final separation and mass transfer rate is consistent with the expectation from angular-momentum conservation: phases of efficient angular-momentum loss, whether through non-conservative mass loss or the onset of a CE, drive stronger orbital decay and produce tighter systems. In our models, close double CO-core WDs tend to descend from such high mass transfer, high angular-momentum loss episodes. No strong correlation is found between WD mass and mass transfer rate.
\end{itemize}

Together, the interconnection between mass transfer rate and orbital separation provides a quantitative illustration of how our adopted mass transfer and angular-momentum loss prescriptions shape the final properties of DWDs, and is consistent with the general expectations from angular-momentum conservation in interacting binaries.

%--------------------------------------------------------------------
\subsection{Dependence on the CE binding parameter}

The characteristics of the DWD population created by our models are determined primarily by the selected CE parameters, specifically by the product  $\alpha_\text{CE}\lambda_\text{CE}$, which controls how much orbital energy is available to unbind the envelope. In our fiducial model, we use $\alpha_\text{CE}=1$ and $\lambda_\text{CE}=0.1$, corresponding to relatively tightly bound envelopes. To assess the robustness of our main outcomes, we perform an additional simulation where we keep $\alpha_\text{CE}=1$ but increase the envelope-structure parameter $\lambda_\text{CE}=0.5$. This choice effectively increases the product $\alpha_\mathrm{CE}\lambda_\mathrm{CE}$, decreases the magnitude of the binding energy of the envelope, allowing a larger fraction of binaries to survive CE ejection and form DWDs. In contrast, the $\lambda_\text{CE}$ parameter does not affect the stability of RLOF mass transfer which instead depends on the donor’s evolutionary stage along with mass ratio, the orbital separation, Roche lobe size, and the adopted mass and angular-momentum prescriptions. Consequently, this model produces a significantly larger number of CE-involving DWDs, while the number of DWDs formed exclusively through the RLOF-only channel remains unchanged relative to the aforementioned fiducial model.

\begin{figure}[htbp] 
    \centering
    \subfigure[~Primary mass.]{\includegraphics[scale=0.49]{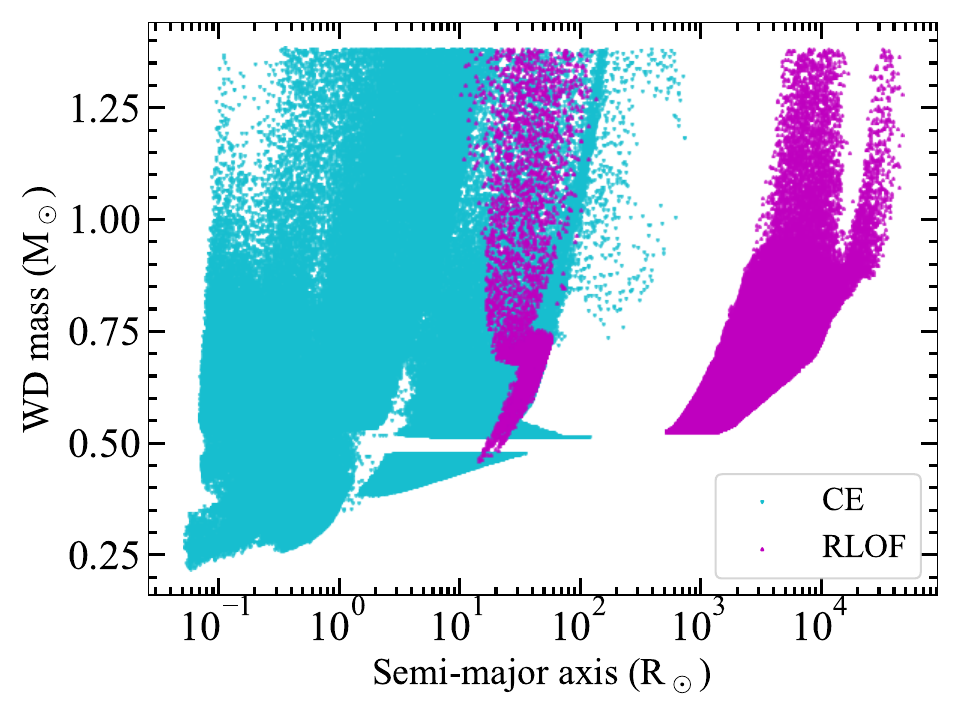}}    
    \subfigure[~Secondary mass.]{\includegraphics[scale=0.49]{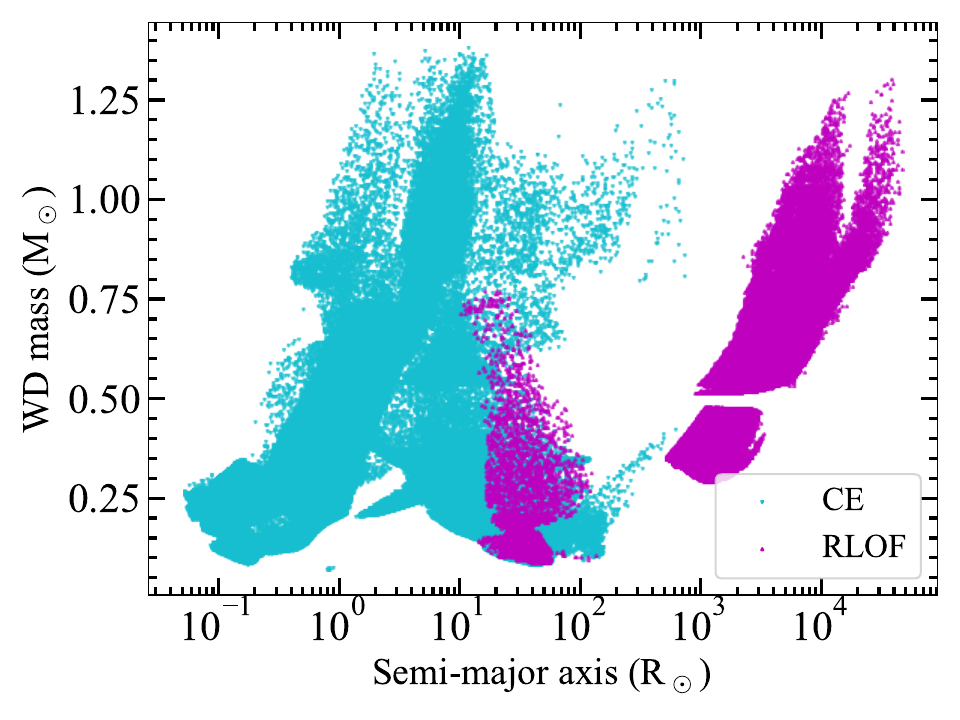}}    
    \caption{Same as Figure~\ref{Fig: CE vs. RLOF} except for $\alpha_\text{CE}=1$ and $\lambda_\text{CE}=0.5$.}
    \label{Fig: CE vs. RLOF2}
\end{figure}

Figure~\ref{Fig: CE vs. RLOF2} illustrates the orbital separations obtained at the end of the RLOF-only and CE-involving channels in the two CE models. The distribution of separations remains clearly bimodal in both cases, with systems that have experienced only RLOF occupying mostly the wide orbits and the CE-involving systems making up the close DWDs. The lack of systems at intermediate separations between tens and a few hundred $\text{R}_\odot$ is intact, although its depth and exact boundaries shift when $\lambda_\text{CE}$ is altered. Increasing $\lambda_\text{CE}$ allows CE survivors to emerge at somewhat larger separations, thus partially filling the gap, which is anticipated from the fact that the orbital shrinkage is lesser when more energy is being used to unbind the envelope.

Overall, the presence of a separation gap in DWD population is a robust feature of our models, while the precise appearance of the gap (its width, depth, and the relative numbers of systems on either side) does depend on the underlying CE parameters. In the remainder of the paper, we therefore interpret the separation gap as a qualitative prediction of our binary-evolution framework, and regard the precise shape of the distribution as a useful diagnostic for future calibration against observed DWD populations.

%--------------------------------------------------------------------
\subsection{Illustrative evolutionary histories}

In order to verify the trends seen in the mass vs. semi-major axis distributions are genuine physical effects and not numerical artifacts, we inspect full evolutionary histories of selected systems drawn from our simulated population. Below, we present two cases that detail the previously mentioned channels.

The first case involves a pair of close DWD, which is composed of a low mass CO WD with a more massive He-core companion. The system originates from an initially unequal mass main sequence binary. The initially more massive star is the first to fill its Roche lobe as a subgiant and undergoes stable mass transfer, being stripped down to a low mass naked He star while while the companion that accretes mass to become the more massive component. At this stage, the orbit widens and circularises. When the companion (now more massive) later ascends the giant branch, its RLOF is dynamically unstable and causes CE to occur. The envelope is ejected, leaving behind a close binary comprising the He star and a newly formed He-core WD. The He star then completes core He burning and cools as a WD. As its stripped core has ignited He and begun to build a CO-core, \textsc{compas} classifies the remnant as a CO-core WD even though its mass is still low. This example demonstrates how early envelope stripping and incomplete He burning naturally yield low mass CO tagged (hybrid He-CO) WDs in close binaries that have come about through a combination of stable RLOF and CE evolution.

The second case is a close and massive DWD composed of a high-mass CO-core WD and a slightly less massive ONe-core star. In this scenario, the star which is initially the less massive becomes a strong mass gainer during an early phase of stable RLOF, growing into a massive stripped He star. After He exhaustion, this star exposes a massive CO-core and cools as a CO WD. The evolution of the companion into a massive giant occurs much later, which results in the initiation of unstable mass transfer leading to a CE phase. The envelope is ejected, and the core of the giant, which has undergone carbon ignition, is left behind as an ONe-core WD in a close orbit around the pre-existing massive CO WD. In this scenario, the CO-core WD is just a bit heavier than its ONe-core partner due to the prior accretion driven growth and the fact that the CO-ONe boundary in \textsc{compas} is determined by core ignition history rather than a sharp mass threshold.

These illustrative full evolutionary histories confirm that the apparently unusual regions of parameter space in our population, such as low mass CO WDs and CO-ONe pairs where the CO component is the more massive star, arise as natural outcomes of well defined mass transfer and CE channels in the binary evolution model. Furthermore, the core type classification method that is used in \textsc{compas} must be taken into account in when analysing the mass and composition distributions of synthetic DWD populations as it is based on the prior burning history of the stripped core.

%--------------------------------------------------------------------
\section{Discussion}\label{Sec4}

In this study, we use the \textsc{compas} simulation to perform a comprehensive population synthesis of how DWDs form and evolve. Our population synthesis results demonstrate that the evolutionary pathway of a binary system shapes the final properties of DWD pairs. A similar low density in the separation distribution of DWDs has been reported previously in combined population synthesis and \textit{Gaia} analyses, and our models reproduce this feature and quantify its dependence on binary interaction physics. The most salient feature in our simulations is the significant bimodal distribution in orbital separation, characterised by a clear gap in $100 - 500\rm\,R_\odot$. This feature emerges directly from the underlying physics of mass transfer: a stable RLOF typically preserves or expands the orbit, whereas CE evolution induces substantial orbital contraction. The way this gap splits the close and wide DWD populations provides a powerful, testable signature for future observatories. GW detectors like \textit{LISA} will be sensitive to the close binary population, while high-precision astrometric surveys (e.g., \textit{Gaia}, \textit{Roman}) can characterise wide binaries, enabling stringent tests of binary evolution models~\citep{2001A&A...368..939N,2012A&A...546A..70T}. In particular, we find that varying the CE binding parameter $\lambda_\text{CE}$ changes the depth and width of the separation gap but not its overall presence, indicating that the gap is a robust outcome of our models while its morphology encodes details of the CE physics.

The core composition of WDs further reflects their evolutionary pathways. Double He-core WDs, with masses $\lesssim0.5\rm\,M_\odot$, in our models form only in close binaries, as they require early envelope stripping via at least one CE phase that also shrinks the orbit. In contrast, CO- and ONe-core WDs are massive and form across a much wider range of orbital separations, reflecting a more diverse set of evolutionary channels~\citep{1985ApJS...58..661I,1998MNRAS.296.1019H,2017A&A...602A..16T}. Notably, our simulation does not produce wide double He-core WD systems. The formation of such a pair would require two consecutive episodes of stable mass transfer that precisely truncate the evolution on RGB for both stars, a scenario demanding a finely tuned set of initial conditions that is statistically improbable. In our simulations such finely tuned configurations do not occur, and we therefore do not find any wide double He-core DWDs. This absence should, however, be interpreted as a consequence of our adopted stability criteria and mass transfer prescriptions rather than as proof that such systems cannot exist in nature.

The observed correlation between the mass transfer rate and final orbital separation offers critical insight into mass transfer physics. High mass transfer rates are associated with brief conservative mass transfer events that result in wide orbits. Conversely, within the population of close binaries, we find that orbital separation decreases with increasing mass transfer rates. Within the population of close binaries we find that systems with smaller final separations tend to have experienced more intense stable mass transfer episodes shortly before DWD formation. This trend is consistent with the expectation that phases of efficient angular-momentum loss during mass transfer, whether through non-conservative mass loss, enhanced winds, or the onset of a CE drive stronger orbital decay and produce tighter DWDs.

One particularly interesting outcome is the substantial population of wide DWDs produced in our simulations. Although their mutual gravitational interaction is weak, these systems remain bound and in principle can serve as natural laboratories for testing fundamental physics. Recent studies incorporating \textit{Gaia} data have shown that wide binaries are excellent probes for modified gravity theories, such as MOND, providing complementary constraints to galaxy dynamics~\citep{2019MNRAS.488.4740P,2024MNRAS.527.4573B}.

Several important assumptions directly affect the properties of the simulated DWD population. The evolution of the CE phase plays a major role: in the standard energy formalism the product $\alpha_\text{CE}\lambda_\text{CE}$ controls how much orbital energy is available to unbind the envelope. Higher values of $\alpha_\text{CE}\lambda_\text{CE}$ make envelope ejection easier, increasing the overall number of CE survivors and favouring somewhat wider post-CE systems, whereas lower values lead to stronger orbital contraction and a higher rate of stellar mergers but also enhance the formation of the tightest DWDs that contribute to the \textit{LISA} band. The stability criterion for RLOF is equally important. A more stringent criterion causes the systems to undergo CE evolution, leading to the formation of tight binaries. In contrast, a more relaxed criterion allows stable RLOF and increases the number of wide binaries. Since the stability boundary is still poorly constrained, variations in this criterion can significantly shift the relative contributions of the RLOF only and CE involving channels in any given population synthesis model. The efficiency of mass transfer ($\beta$, the fraction of transferred mass that is accreted by the companion) also influences the outcomes. In many commonly used prescriptions, nearly conservative transfer ($\beta = 1$) tends to favour orbital expansion for mass ratios close to unity, whereas non-conservative transfer ($\beta<1$) can remove substantial orbital angular momentum through ejected material and often leads to orbital contraction, especially for unequal mass systems.

In addition to these binary interaction parameters, outputs are also affected by global parameters including the metallicity and the star formation rate (SFR). Lower metallicity environments typically form more massive WDs and alter CE survival rates, whereas higher metallicity favours low mass remnants and wider systems. Similarly, a constant SFR assumption provides a time averaged view of the Galactic population, but a bursty or declining SFR would change the relative proportions of young, close binaries to older, wider binaries. For simplicity, we fixed the metallicity and SFR in this work, and a detailed systematic study of their effects is reserved for future work. Nonetheless, within the range of parameter variations we have explored, the qualitative features of the population, such as the presence of a separation gap between RLOF only and CE involving systems and the broad mass and core type trends remain robust, even though the detailed morphology and relative weights of the different channels can change significantly.

%--------------------------------------------------------------------
\subsection{DWD formation efficiency and comparison with previous studies}

In our fiducial \textsc{compas} run ($\alpha_\text{CE}\lambda_\text{CE}=0.1$), we evolve $N_{\rm bin} = 10^{7}$ ZAMS binaries with $Z=0.014$ and a total initial stellar mass $M_{\rm init,tot} = 2.021045\times 10^{7}\rm\,M_\odot$. Among them, $N_{\rm DWD} = 329\,147$ systems form DWDs, corresponding to a formation efficiency
\begin{equation}
f_{\rm DWD} = \frac{N_{\rm DWD}}{N_{\rm bin}} \simeq 3.29\times 10^{-2},
\end{equation}
and a DWD yield per unit stellar mass formed
\begin{equation}
\eta_{\rm DWD} = \frac{N_{\rm DWD}}{M_{\rm init,tot}} \simeq 1.63\times 10^{-2}\rm\,M_\odot^{-1}.
\end{equation}
Breaking this down by interaction channel, $N_{\rm DWD,RLOF} = 176\,915$ systems ($\simeq 54\%$) form through only stable RLOF, while $N_{\rm DWD,CE} = 115\,182$ systems ($\simeq 35\%$) experience at least one CE episode. The residual $N_{\rm DWD,other} = 37\,050$ systems ($\simeq 11\%$) form without any mass transfer episode.

Assuming the birth rate of DWDs ($\mathcal{R}_{\rm DWD}$) scales with the SFR of Milky Way (MW), it can be expressed as
\begin{equation}
\mathcal{R}_{\rm DWD} = \eta_{\rm DWD}\times\text{SFR}_\text{MW}.
\end{equation}
Following the recent estimate by \cite{2023A&A...672A..54S}, we choose $\text{SFR}_\text{MW} = 6\pm2 \rm\, M_\odot\,yr^{-1}$, which yields $\mathcal{R}_{\rm DWD} \simeq (9.6\pm3.2)\times 10^{-2}\,{\rm yr}^{-1}$. With the channel splits, these values turn out to be $\mathcal{R}_{\rm DWD,RLOF}\simeq (5.28\pm1.76)\times 10^{-2} \,{\rm yr}^{-1}$, $\mathcal{R}_{\rm DWD,CE} \simeq (3.42\pm1.14)\times 10^{-2} \,{\rm yr}^{-1}$, and $\mathcal{R}_{\rm DWD,other} \simeq (1.08\pm0.36)\times 10^{-2} \,{\rm yr}^{-1}$. In the second simulation with $\alpha_\text{CE}\lambda_\text{CE}=0.5$, these numbers change to $\mathcal{R}_{\rm DWD,RLOF}\simeq (5.28\pm1.76)\times 10^{-2} \,{\rm yr}^{-1}$, $\mathcal{R}_{\rm DWD,CE} \simeq (9.47\pm3.16)\times 10^{-2} \,{\rm yr}^{-1}$, and $\mathcal{R}_{\rm DWD,other} \simeq (1.08\pm0.36)\times 10^{-2} \,{\rm yr}^{-1}$. This simple scaling can be compared with earlier Galactic DWD population synthesis studies. Using \textsc{SeBa} binary population synthesis code, \cite{2001A&A...365..491N} reported a Galactic birth rate of close DWDs of $\simeq 5\times 10^{-2}\,{\rm yr}^{-1}$ with a total present day population of $\simeq 2.5\times 10^{8}$ systems. Other rapid binary-evolution models that adopt the prescriptions of \cite{2002MNRAS.329..897H}, typically predict current DWD birth rates of $3.21\times 10^{-2}\,{\rm yr}^{-1}$ in the thin disc, depending on the assumed star-formation history \citep{2010A&A...521A..85Y}. Despite subtle differences in input physics and calibration, our \textsc{compas}-based rates are thus broadly consistent with previous population synthesis results obtained using independent codes and assumptions within the expected factor of few systematic uncertainties.

%--------------------------------------------------------------------
\section{Conclusions}\label{Sec5}

In this work, we have used the rapid binary population synthesis code \textsc{compas} to explore how evolutionary pathways, WD core compositions, and mass transfer rates jointly shape the final properties of DWD systems. Our models reproduce the bimodal separation distribution and intermediate under-density previously identified in studies combining population synthesis with \textit{Gaia} data. Thereby we show that this feature arises naturally from the coexistence of two broad families of interaction channels: wide systems that evolve purely through stable RLOF and close systems that experience at least one CE phase. Within this unified framework we quantify how the positions of DWDs in the mass--separation plane correlate with their core types (He, CO, ONe) and with the intensity of their last stable mass transfer episode, and we demonstrate that the presence of a separation gap between wide RLOF only and close CE-involving systems is robust to reasonable changes in the CE structure parameter, even though its detailed morphology does depend on the adopted CE physics.

These results provide a physically interpretable map between the observed parameters of DWDs (masses, separations, and core types) and the underlying interaction histories predicted by \textsc{compas} like binary evolution modes. They therefore offer a concrete framework for using future observations to calibrate uncertain aspects of binary physics, such as the stability of mass transfer and the efficiency of CE ejection. Looking ahead, forthcoming observatories will greatly improve our empirical picture of the DWD population: space-based GW missions such as \textit{LISA} will systematically survey the Galactic population of close, interacting binaries, while astrometric missions such as \textit{Gaia} and \textit{Roman} will enhance the detection and characterisation of wide DWDs. Combining these data sets with population synthesis models like those presented here will be crucial for refining binary evolution prescriptions and may eventually allow tests of gravitational physics on binary scales.

\begin{acknowledgements}
      The authors would like to thank the anonymous reviewer for their insightful comments, which have helped to improve the manuscript content. We thank T. Bulik and D. Rosi\'nska from University of Warsaw for useful discussion during the compilation of the work. SK acknowledge funding from the National Center for Science, Poland (grant no. 2023/49/B/ST9/02777).
\end{acknowledgements}

\section*{Data Availability}

The data used in this article are obtained from the \textsc{compas} v02.50.00 simulation, which will be shared on reasonable request to the corresponding author.

\bibliographystyle{paslike}
\bibliography{v1}

@ARTICLE{2001A&A...375..890N,
       author = {{Nelemans}, G. and {Yungelson}, L.~R. and {Portegies Zwart}, S.~F.},
        title = "{The gravitational wave signal from the Galactic disk population of binaries containing two compact objects}",
      journal = {\aap},
         year = 2001,
        month = sep,
       volume = {375},
        pages = {890-898},
          doi = {10.1051/0004-6361:20010683},
archivePrefix = {arXiv},
       eprint = {astro-ph/0105221},
 primaryClass = {astro-ph},
       adsurl = {https://ui.adsabs.harvard.edu/abs/2001A&A...375..890N}
}

@ARTICLE{2020ApJ...898...71B,
       author = {{Breivik}, Katelyn and {Coughlin}, Scott and {Zevin}, Michael and {Rodriguez}, Carl L. and {Kremer}, Kyle and {Ye}, Claire S. and {Andrews}, Jeff J. and {Kurkowski}, Michael and {Digman}, Matthew C. and {Larson}, Shane L. and {Rasio}, Frederic A.},
        title = "{COSMIC Variance in Binary Population Synthesis}",
      journal = {\apj},
         year = 2020,
        month = jul,
       volume = {898},
       number = {1},
          eid = {71},
        pages = {71},
          doi = {10.3847/1538-4357/ab9d85},
archivePrefix = {arXiv},
       eprint = {1911.00903},
 primaryClass = {astro-ph.HE},
       adsurl = {https://ui.adsabs.harvard.edu/abs/2020ApJ...898...71B}
}

@ARTICLE{2012A&A...546A..70T,
       author = {{Toonen}, S. and {Nelemans}, G. and {Portegies Zwart}, S.},
        title = "{Supernova Type Ia progenitors from merging double white dwarfs. Using a new population synthesis model}",
      journal = {\aap},
         year = 2012,
        month = oct,
       volume = {546},
          eid = {A70},
        pages = {A70},
          doi = {10.1051/0004-6361/201218966},
archivePrefix = {arXiv},
       eprint = {1208.6446},
 primaryClass = {astro-ph.HE},
       adsurl = {https://ui.adsabs.harvard.edu/abs/2012A&A...546A..70T}
}

@ARTICLE{2017A&A...602A..16T,
       author = {{Toonen}, S. and {Hollands}, M. and {G{\"a}nsicke}, B.~T. and {Boekholt}, T.},
        title = "{The binarity of the local white dwarf population}",
      journal = {\aap},
         year = 2017,
        month = jun,
       volume = {602},
          eid = {A16},
        pages = {A16},
          doi = {10.1051/0004-6361/201629978},
archivePrefix = {arXiv},
       eprint = {1703.06893},
 primaryClass = {astro-ph.SR},
       adsurl = {https://ui.adsabs.harvard.edu/abs/2017A&A...602A..16T}
}

@ARTICLE{2001PASP..113..409F,
       author = {{Fontaine}, G. and {Brassard}, P. and {Bergeron}, P.},
        title = "{The Potential of White Dwarf Cosmochronology}",
      journal = {\pasp},
         year = 2001,
        month = apr,
       volume = {113},
       number = {782},
        pages = {409-435},
          doi = {10.1086/319535},
       adsurl = {https://ui.adsabs.harvard.edu/abs/2001PASP..113..409F}
}

@ARTICLE{2010A&ARv..18..471A,
       author = {{Althaus}, Leandro G. and {C{\'o}rsico}, Alejandro H. and {Isern}, Jordi and {Garc{\'\i}a-Berro}, Enrique},
        title = "{Evolutionary and pulsational properties of white dwarf stars}",
      journal = {\aapr},
         year = 2010,
        month = oct,
       volume = {18},
       number = {4},
        pages = {471-566},
          doi = {10.1007/s00159-010-0033-1},
archivePrefix = {arXiv},
       eprint = {1007.2659},
 primaryClass = {astro-ph.SR},
       adsurl = {https://ui.adsabs.harvard.edu/abs/2010A&ARv..18..471A}
}

@ARTICLE{1984ApJ...277..355W,
       author = {{Webbink}, R.~F.},
        title = "{Double white dwarfs as progenitors of R Coronae Borealis stars and type I supernovae.}",
      journal = {\apj},
         year = 1984,
        month = feb,
       volume = {277},
        pages = {355-360},
          doi = {10.1086/161701},
       adsurl = {https://ui.adsabs.harvard.edu/abs/1984ApJ...277..355W}
}

@ARTICLE{1997ApJ...475..291I,
       author = {{Iben}, Jr., Icko and {Tutukov}, Alexander V. and {Yungelson}, Lev R.},
        title = "{Helium and Carbon-Oxygen White Dwarfs in Close Binaries}",
      journal = {\apj},
         year = 1997,
        month = jan,
       volume = {475},
       number = {1},
        pages = {291-299},
          doi = {10.1086/303525},
       adsurl = {https://ui.adsabs.harvard.edu/abs/1997ApJ...475..291I}
}

@ARTICLE{1985ApJS...58..661I,
       author = {{Iben}, Jr., I. and {Tutukov}, A.~V.},
        title = "{On the evolution of close binaries with components of initial mass between 3 M and 12 M.}",
      journal = {\apjs},
         year = 1985,
        month = aug,
       volume = {58},
        pages = {661-710},
          doi = {10.1086/191054},
       adsurl = {https://ui.adsabs.harvard.edu/abs/1985ApJS...58..661I}
}

@ARTICLE{1987ApJ...323..129E,
       author = {{Evans}, Charles R. and {Iben}, Jr., Icko and {Smarr}, Larry},
        title = "{Degenerate Dwarf Binaries as Promising, Detectable Sources of Gravitational Radiation}",
      journal = {\apj},
         year = 1987,
        month = dec,
       volume = {323},
        pages = {129},
          doi = {10.1086/165812},
       adsurl = {https://ui.adsabs.harvard.edu/abs/1987ApJ...323..129E}
}

@ARTICLE{1986ApJ...311..753I,
       author = {{Iben}, Jr., Icko and {Tutukov}, Alexander V.},
        title = "{On the Number-Mass Distribution of Degenerate Dwarfs Produced by the Interacting Binaries and Evidence for Mergers of Low-Mass Helium Dwarfs}",
      journal = {\apj},
         year = 1986,
        month = dec,
       volume = {311},
        pages = {753},
          doi = {10.1086/164813},
       adsurl = {https://ui.adsabs.harvard.edu/abs/1986ApJ...311..753I}
}

@ARTICLE{1987ApJ...313..727I,
       author = {{Iben}, Jr., Icko and {Tutukov}, Alexander V.},
        title = "{Evolutionary Scenarios for Intermediate-Mass Stars in Close Binaries}",
      journal = {\apj},
         year = 1987,
        month = feb,
       volume = {313},
        pages = {727},
          doi = {10.1086/165011},
       adsurl = {https://ui.adsabs.harvard.edu/abs/1987ApJ...313..727I}
}

@ARTICLE{1988Ap&SS.145....1L,
       author = {{Lipunov}, V.~M. and {Postnov}, K.~A.},
        title = "{The Joint Evolution of Normal and Compact Magnetized Stars in Close Binaries - Analytical Description and Statistical Simulation}",
      journal = {\apss},
         year = 1988,
        month = jun,
       volume = {145},
       number = {1},
        pages = {1-45},
          doi = {10.1007/BF00645692},
       adsurl = {https://ui.adsabs.harvard.edu/abs/1988Ap&SS.145....1L}
}

@INPROCEEDINGS{1993ASPC...38..211Y,
       author = {{Yungelson}, L.~R. and {Tutukov}, A.~V.},
        title = "{Model of the Galactical Population of White Dwarfs}",
    booktitle = {New Frontiers in Binary Star Research},
         year = 1993,
       editor = {{Leung}, Kam-Ching and {Nha}, Il-Seong},
       series = {Astronomical Society of the Pacific Conference Series},
       volume = {38},
        month = jan,
        pages = {211},
       adsurl = {https://ui.adsabs.harvard.edu/abs/1993ASPC...38..211Y}
}

@ARTICLE{1995MNRAS.272..800H,
       author = {{Han}, Zhanwen and {Podsiadlowski}, Philipp and {Eggleton}, Peter P.},
        title = "{The formation of bipolar planetary nebulae and close white dwarf binaries}",
      journal = {\mnras},
         year = 1995,
        month = feb,
       volume = {272},
       number = {4},
        pages = {800-820},
          doi = {10.1093/mnras/272.4.800},
       adsurl = {https://ui.adsabs.harvard.edu/abs/1995MNRAS.272..800H}
}

@ARTICLE{1998MNRAS.296.1019H,
       author = {{Han}, Zhanwen},
        title = "{The formation of double degenerates and related objects}",
      journal = {\mnras},
         year = 1998,
        month = jun,
       volume = {296},
       number = {4},
        pages = {1019-1040},
          doi = {10.1046/j.1365-8711.1998.01475.x},
       adsurl = {https://ui.adsabs.harvard.edu/abs/1998MNRAS.296.1019H}
}

@ARTICLE{2020ApJ...889...49B,
       author = {{Brown}, Warren R. and {Kilic}, Mukremin and {Kosakowski}, Alekzander and {Andrews}, Jeff J. and {Heinke}, Craig O. and {Ag{\"u}eros}, Marcel A. and {Camilo}, Fernando and {Gianninas}, A. and {Hermes}, J.~J. and {Kenyon}, Scott J.},
        title = "{The ELM Survey. VIII. Ninety-eight Double White Dwarf Binaries}",
      journal = {\apj},
         year = 2020,
        month = jan,
       volume = {889},
       number = {1},
          eid = {49},
        pages = {49},
          doi = {10.3847/1538-4357/ab63cd},
archivePrefix = {arXiv},
       eprint = {2002.00064},
 primaryClass = {astro-ph.SR},
       adsurl = {https://ui.adsabs.harvard.edu/abs/2020ApJ...889...49B}
}

@ARTICLE{2022MNRAS.511.5936K,
       author = {{Korol}, Valeriya and {Hallakoun}, Na'ama and {Toonen}, Silvia and {Karnesis}, Nikolaos},
        title = "{Observationally driven Galactic double white dwarf population for LISA}",
      journal = {\mnras},
         year = 2022,
        month = apr,
       volume = {511},
       number = {4},
        pages = {5936-5947},
          doi = {10.1093/mnras/stac415},
archivePrefix = {arXiv},
       eprint = {2109.10972},
 primaryClass = {astro-ph.HE},
       adsurl = {https://ui.adsabs.harvard.edu/abs/2022MNRAS.511.5936K}
}

@ARTICLE{1983ApJ...268..368E,
       author = {{Eggleton}, P.~P.},
        title = "{Aproximations to the radii of Roche lobes.}",
      journal = {\apj},
         year = 1983,
        month = may,
       volume = {268},
        pages = {368-369},
          doi = {10.1086/160960},
       adsurl = {https://ui.adsabs.harvard.edu/abs/1983ApJ...268..368E}
}

@INPROCEEDINGS{1976IAUS...73...75P,
       author = {{Paczynski}, B.},
        title = "{Common Envelope Binaries}",
    booktitle = {Structure and Evolution of Close Binary Systems},
         year = 1976,
       editor = {{Eggleton}, Peter and {Mitton}, Simon and {Whelan}, John},
       series = {IAU Symposium},
       volume = {73},
        month = jan,
        pages = {75},
       adsurl = {https://ui.adsabs.harvard.edu/abs/1976IAUS...73...75P}
}

@BOOK{1985ibs..book.....P,
       author = {{Pringle}, J.~E. and {Wade}, R.~A.},
        title = "{Interacting binary stars}",
    publisher = "{Cambridge University Press}",
         year = 1985,
       adsurl = {https://ui.adsabs.harvard.edu/abs/1985ibs..book.....P}
}

@PROCEEDINGS{1996ASIC..477.....W,
        title = "{Evolutionary processes in binary stars}",
    booktitle = {Evolutionary Processes in Binary Stars},
         year = 1996,
       editor = {{Wijers}, Ralph A.~M.~J. and {Davies}, Melvyn B. and {Tout}, Christopher A.},
       series = {NATO Advanced Study Institute (ASI) Series C},
       volume = {477},
        month = jan,
       adsurl = {https://ui.adsabs.harvard.edu/abs/1996ASIC..477.....W}
}

@INPROCEEDINGS{1979IAUS...83..401T,
       author = {{Tutukov}, A. and {Yungelson}, L.},
        title = "{Evolution of massive common envelope binaries and mass loss.}",
    booktitle = {Mass Loss and Evolution of O-Type Stars},
         year = 1979,
       editor = {{Conti}, P.~S. and {De Loore}, C.~W.~H.},
       series = {IAU Symposium},
       volume = {83},
        month = jan,
        pages = {401-406},
       adsurl = {https://ui.adsabs.harvard.edu/abs/1979IAUS...83..401T}
}

@ARTICLE{1993PASP..105.1373I,
       author = {{Iben}, Jr., Icko and {Livio}, Mario},
        title = "{Common Envelopes in Binary Star Evolution}",
      journal = {\pasp},
         year = 1993,
        month = dec,
       volume = {105},
        pages = {1373},
          doi = {10.1086/133321},
       adsurl = {https://ui.adsabs.harvard.edu/abs/1993PASP..105.1373I}
}

@ARTICLE{2022ApJS..258...34R,
       author = {{Riley}, Jeff and {Agrawal}, Poojan and {Barrett}, Jim W. and {Boyett}, Kristan N.~K. and {Broekgaarden}, Floor S. and {Chattopadhyay}, Debatri and {Gaebel}, Sebastian M. and {Gittins}, Fabian and {Hirai}, Ryosuke and {Howitt}, George and {Justham}, Stephen and {Khandelwal}, Lokesh and {Kummer}, Floris and {Lau}, Mike Y.~M. and {Mandel}, Ilya and {de Mink}, Selma E. and {Neijssel}, Coenraad and {Riley}, Tim and {van Son}, Lieke and {Stevenson}, Simon and {Vigna-G{\'o}mez}, Alejandro and {Vinciguerra}, Serena and {Wagg}, Tom and {Willcox}, Reinhold and {Team Compas}},
        title = "{Rapid Stellar and Binary Population Synthesis with COMPAS}",
      journal = {\apjs},
         year = 2022,
        month = feb,
       volume = {258},
       number = {2},
          eid = {34},
        pages = {34},
          doi = {10.3847/1538-4365/ac416c},
archivePrefix = {arXiv},
       eprint = {2109.10352},
 primaryClass = {astro-ph.IM},
       adsurl = {https://ui.adsabs.harvard.edu/abs/2022ApJS..258...34R}
}

@ARTICLE{2025ApJS..280...43T,
       author = {{Team COMPAS} and {Mandel}, Ilya and {Riley}, Jeff and {Boesky}, Adam and {Br{\v{c}}ek}, Adam and {Hirai}, Ryosuke and {Kapil}, Veome and {Lau}, Mike Y.~M. and {Merritt}, Jd and {Rodr{\'\i}guez-Segovia}, Nicol{\'a}s and {Romero-Shaw}, Isobel and {Song}, Yuzhe and {Stevenson}, Simon and {Vajpeyi}, Avi and {van Son}, L.~A.~C. and {Vigna-G{\'o}mez}, Alejandro and {Willcox}, Reinhold},
        title = "{Rapid Stellar and Binary Population Synthesis with COMPAS: Methods Paper II}",
      journal = {\apjs},
         year = 2025,
        month = sep,
       volume = {280},
       number = {1},
          eid = {43},
        pages = {43},
          doi = {10.3847/1538-4365/adf8d0},
archivePrefix = {arXiv},
       eprint = {2506.02316},
 primaryClass = {astro-ph.SR},
       adsurl = {https://ui.adsabs.harvard.edu/abs/2025ApJS..280...43T}
}

@ARTICLE{2002MNRAS.329..897H,
       author = {{Hurley}, Jarrod R. and {Tout}, Christopher A. and {Pols}, Onno R.},
        title = "{Evolution of binary stars and the effect of tides on binary populations}",
      journal = {\mnras},
         year = 2002,
        month = feb,
       volume = {329},
       number = {4},
        pages = {897-928},
          doi = {10.1046/j.1365-8711.2002.05038.x},
archivePrefix = {arXiv},
       eprint = {astro-ph/0201220},
 primaryClass = {astro-ph},
       adsurl = {https://ui.adsabs.harvard.edu/abs/2002MNRAS.329..897H}
}

@ARTICLE{2008ApJS..174..223B,
       author = {{Belczynski}, Krzysztof and {Kalogera}, Vassiliki and {Rasio}, Frederic A. and {Taam}, Ronald E. and {Zezas}, Andreas and {Bulik}, Tomasz and {Maccarone}, Thomas J. and {Ivanova}, Natalia},
        title = "{Compact Object Modeling with the StarTrack Population Synthesis Code}",
      journal = {\apjs},
         year = 2008,
        month = jan,
       volume = {174},
       number = {1},
        pages = {223-260},
          doi = {10.1086/521026},
archivePrefix = {arXiv},
       eprint = {astro-ph/0511811},
 primaryClass = {astro-ph},
       adsurl = {https://ui.adsabs.harvard.edu/abs/2008ApJS..174..223B}
}

@ARTICLE{2004MNRAS.350..407I,
       author = {{Izzard}, Robert G. and {Tout}, Christopher A. and {Karakas}, Amanda I. and {Pols}, Onno R.},
        title = "{A new synthetic model for asymptotic giant branch stars}",
      journal = {\mnras},
         year = 2004,
        month = may,
       volume = {350},
       number = {2},
        pages = {407-426},
          doi = {10.1111/j.1365-2966.2004.07446.x},
archivePrefix = {arXiv},
       eprint = {astro-ph/0402403},
 primaryClass = {astro-ph},
       adsurl = {https://ui.adsabs.harvard.edu/abs/2004MNRAS.350..407I}
}

@ARTICLE{2018PASA...35...31P,
       author = {{Price}, Daniel J. and {Wurster}, James and {Tricco}, Terrence S. and {Nixon}, Chris and {Toupin}, St{\'e}ven and {Pettitt}, Alex and {Chan}, Conrad and {Mentiplay}, Daniel and {Laibe}, Guillaume and {Glover}, Simon and {Dobbs}, Clare and {Nealon}, Rebecca and {Liptai}, David and {Worpel}, Hauke and {Bonnerot}, Cl{\'e}ment and {Dipierro}, Giovanni and {Ballabio}, Giulia and {Ragusa}, Enrico and {Federrath}, Christoph and {Iaconi}, Roberto and {Reichardt}, Thomas and {Forgan}, Duncan and {Hutchison}, Mark and {Constantino}, Thomas and {Ayliffe}, Ben and {Hirsh}, Kieran and {Lodato}, Giuseppe},
        title = "{Phantom: A Smoothed Particle Hydrodynamics and Magnetohydrodynamics Code for Astrophysics}",
      journal = {\pasa},
         year = 2018,
        month = sep,
       volume = {35},
          eid = {e031},
        pages = {e031},
          doi = {10.1017/pasa.2018.25},
archivePrefix = {arXiv},
       eprint = {1702.03930},
 primaryClass = {astro-ph.IM},
       adsurl = {https://ui.adsabs.harvard.edu/abs/2018PASA...35...31P}
}

@ARTICLE{2022MNRAS.516.5737B,
       author = {{Broekgaarden}, Floor S. and {Berger}, Edo and {Stevenson}, Simon and {Justham}, Stephen and {Mandel}, Ilya and {Chru{\'s}li{\'n}ska}, Martyna and {van Son}, Lieke A.~C. and {Wagg}, Tom and {Vigna-G{\'o}mez}, Alejandro and {de Mink}, Selma E. and {Chattopadhyay}, Debatri and {Neijssel}, Coenraad J.},
        title = "{Impact of massive binary star and cosmic evolution on gravitational wave observations - II. Double compact object rates and properties}",
      journal = {\mnras},
         year = 2022,
        month = nov,
       volume = {516},
       number = {4},
        pages = {5737-5761},
          doi = {10.1093/mnras/stac1677},
archivePrefix = {arXiv},
       eprint = {2112.05763},
 primaryClass = {astro-ph.HE},
       adsurl = {https://ui.adsabs.harvard.edu/abs/2022MNRAS.516.5737B}
}

@ARTICLE{2022MNRAS.517.4034S,
       author = {{Stevenson}, Simon and {Clarke}, Teagan A.},
        title = "{Constraints on the contributions to the observed binary black hole population from individual evolutionary pathways in isolated binary evolution}",
      journal = {\mnras},
         year = 2022,
        month = dec,
       volume = {517},
       number = {3},
        pages = {4034-4053},
          doi = {10.1093/mnras/stac2936},
archivePrefix = {arXiv},
       eprint = {2210.05040},
 primaryClass = {astro-ph.HE},
       adsurl = {https://ui.adsabs.harvard.edu/abs/2022MNRAS.517.4034S}
}

@ARTICLE{2022ApJ...937..118W,
       author = {{Wagg}, T. and {Broekgaarden}, F.~S. and {de Mink}, S.~E. and {Frankel}, N. and {van Son}, L.~A.~C. and {Justham}, S.},
        title = "{Gravitational Wave Sources in Our Galactic Backyard: Predictions for BHBH, BHNS, and NSNS Binaries Detectable with LISA}",
      journal = {\apj},
         year = 2022,
        month = oct,
       volume = {937},
       number = {2},
          eid = {118},
        pages = {118},
          doi = {10.3847/1538-4357/ac8675},
archivePrefix = {arXiv},
       eprint = {2111.13704},
 primaryClass = {astro-ph.HE},
       adsurl = {https://ui.adsabs.harvard.edu/abs/2022ApJ...937..118W}
}

@ARTICLE{2010ApJ...710.1310M,
       author = {{Meng}, X. and {Yang}, W.},
        title = "{A Comprehensive Progenitor Model for SNe Ia}",
      journal = {\apj},
         year = 2010,
        month = feb,
       volume = {710},
       number = {2},
        pages = {1310-1323},
          doi = {10.1088/0004-637X/710/2/1310},
archivePrefix = {arXiv},
       eprint = {0910.4992},
 primaryClass = {astro-ph.SR},
       adsurl = {https://ui.adsabs.harvard.edu/abs/2010ApJ...710.1310M}
}

@ARTICLE{2015ApJ...807..105S,
       author = {{Sato}, Yushi and {Nakasato}, Naohito and {Tanikawa}, Ataru and {Nomoto}, Ken'ichi and {Maeda}, Keiichi and {Hachisu}, Izumi},
        title = "{A Systematic Study of Carbon-Oxygen White Dwarf Mergers: Mass Combinations for Type Ia Supernovae}",
      journal = {\apj},
         year = 2015,
        month = jul,
       volume = {807},
       number = {1},
          eid = {105},
        pages = {105},
          doi = {10.1088/0004-637X/807/1/105},
archivePrefix = {arXiv},
       eprint = {1505.01646},
 primaryClass = {astro-ph.HE},
       adsurl = {https://ui.adsabs.harvard.edu/abs/2015ApJ...807..105S}
}

@ARTICLE{2018MNRAS.480.4519C,
       author = {{Canals}, Pere and {Torres}, Santiago and {Soker}, Noam},
        title = "{Oxygen-neon-rich merger during common envelope evolution}",
      journal = {\mnras},
         year = 2018,
        month = nov,
       volume = {480},
       number = {4},
        pages = {4519-4525},
          doi = {10.1093/mnras/sty2121},
archivePrefix = {arXiv},
       eprint = {1806.06730},
 primaryClass = {astro-ph.SR},
       adsurl = {https://ui.adsabs.harvard.edu/abs/2018MNRAS.480.4519C}
}

@ARTICLE{2019MNRAS.490.5888L,
       author = {{Lamberts}, Astrid and {Blunt}, Sarah and {Littenberg}, Tyson B. and {Garrison-Kimmel}, Shea and {Kupfer}, Thomas and {Sanderson}, Robyn E.},
        title = "{Predicting the LISA white dwarf binary population in the Milky Way with cosmological simulations}",
      journal = {\mnras},
         year = 2019,
        month = dec,
       volume = {490},
       number = {4},
        pages = {5888-5903},
          doi = {10.1093/mnras/stz2834},
archivePrefix = {arXiv},
       eprint = {1907.00014},
 primaryClass = {astro-ph.HE},
       adsurl = {https://ui.adsabs.harvard.edu/abs/2019MNRAS.490.5888L}
}

@ARTICLE{2024A&A...686A.221R,
       author = {{Rebassa-Mansergas}, Alberto and {Hollands}, Mark and {Parsons}, Steven G. and {Althaus}, Leandro G. and {Pelisoli}, Ingrid and {Irawati}, Puji and {Raddi}, Roberto and {Camisassa}, Maria E. and {Torres}, Santiago},
        title = "{J0526+5934: A peculiar ultra-short-period double white dwarf}",
      journal = {\aap},
         year = 2024,
        month = jun,
       volume = {686},
          eid = {A221},
        pages = {A221},
          doi = {10.1051/0004-6361/202449519},
archivePrefix = {arXiv},
       eprint = {2402.04443},
 primaryClass = {astro-ph.SR},
       adsurl = {https://ui.adsabs.harvard.edu/abs/2024A&A...686A.221R}
}

@ARTICLE{2009ApJ...697.1048P,
       author = {{Perets}, Hagai B. and {Fabrycky}, Daniel C.},
        title = "{On the Triple Origin of Blue Stragglers}",
      journal = {\apj},
         year = 2009,
        month = jun,
       volume = {697},
       number = {2},
        pages = {1048-1056},
          doi = {10.1088/0004-637X/697/2/1048},
archivePrefix = {arXiv},
       eprint = {0901.4328},
 primaryClass = {astro-ph.SR},
       adsurl = {https://ui.adsabs.harvard.edu/abs/2009ApJ...697.1048P}
}

@ARTICLE{2012ApJ...744...12W,
       author = {{Woods}, T.~E. and {Ivanova}, N. and {van der Sluys}, M.~V. and {Chaichenets}, S.},
        title = "{On the Formation of Double White Dwarfs through Stable Mass Transfer and a Common Envelope}",
      journal = {\apj},
         year = 2012,
        month = jan,
       volume = {744},
       number = {1},
          eid = {12},
        pages = {12},
          doi = {10.1088/0004-637X/744/1/12},
archivePrefix = {arXiv},
       eprint = {1102.1039},
 primaryClass = {astro-ph.SR},
       adsurl = {https://ui.adsabs.harvard.edu/abs/2012ApJ...744...12W}
}

@ARTICLE{2023MNRAS.518.3966S,
       author = {{Scherbak}, Peter and {Fuller}, Jim},
        title = "{White dwarf binaries suggest a common envelope efficiency {\ensuremath{\alpha}}   1/3}",
      journal = {\mnras},
         year = 2023,
        month = jan,
       volume = {518},
       number = {3},
        pages = {3966-3984},
          doi = {10.1093/mnras/stac3313},
archivePrefix = {arXiv},
       eprint = {2211.02036},
 primaryClass = {astro-ph.SR},
       adsurl = {https://ui.adsabs.harvard.edu/abs/2023MNRAS.518.3966S}
}

@ARTICLE{2024A&A...692A.165T,
       author = {{Toubiana}, A. and {Karnesis}, N. and {Lamberts}, A. and {Miller}, M.~C.},
        title = "{The interacting double white dwarf population with LISA: Stochastic foreground and resolved sources}",
      journal = {\aap},
         year = 2024,
        month = dec,
       volume = {692},
          eid = {A165},
        pages = {A165},
          doi = {10.1051/0004-6361/202450174},
archivePrefix = {arXiv},
       eprint = {2403.16867},
 primaryClass = {astro-ph.SR},
       adsurl = {https://ui.adsabs.harvard.edu/abs/2024A&A...692A.165T}
}

@ARTICLE{2025A&A...699A.172V,
       author = {{van Zeist}, Wouter G.~J. and {van Roestel}, Jan and {Nelemans}, Gijs and {Eldridge}, Jan J. and {Korol}, Valeriya and {Toonen}, Silvia},
        title = "{Comparing population synthesis models of compact double white dwarfs to electromagnetic observations}",
      journal = {\aap},
         year = 2025,
        month = jul,
       volume = {699},
          eid = {A172},
        pages = {A172},
          doi = {10.1051/0004-6361/202554302},
archivePrefix = {arXiv},
       eprint = {2505.20953},
 primaryClass = {astro-ph.SR},
       adsurl = {https://ui.adsabs.harvard.edu/abs/2025A&A...699A.172V}
}

@ARTICLE{1975MNRAS.173..729H,
       author = {{Heggie}, D.~C.},
        title = "{Binary evolution in stellar dynamics.}",
      journal = {\mnras},
         year = 1975,
        month = dec,
       volume = {173},
        pages = {729-787},
          doi = {10.1093/mnras/173.3.729},
       adsurl = {https://ui.adsabs.harvard.edu/abs/1975MNRAS.173..729H}
}

@ARTICLE{1955ApJ...121..161S,
       author = {{Salpeter}, Edwin E.},
        title = "{The Luminosity Function and Stellar Evolution.}",
      journal = {\apj},
         year = 1955,
        month = jan,
       volume = {121},
        pages = {161},
          doi = {10.1086/145971},
       adsurl = {https://ui.adsabs.harvard.edu/abs/1955ApJ...121..161S}
}

@ARTICLE{2013A&ARv..21...59I,
       author = {{Ivanova}, N. and {Justham}, S. and {Chen}, X. and {De Marco}, O. and {Fryer}, C.~L. and {Gaburov}, E. and {Ge}, H. and {Glebbeek}, E. and {Han}, Z. and {Li}, X. -D. and {Lu}, G. and {Marsh}, T. and {Podsiadlowski}, P. and {Potter}, A. and {Soker}, N. and {Taam}, R. and {Tauris}, T.~M. and {van den Heuvel}, E.~P.~J. and {Webbink}, R.~F.},
        title = "{Common envelope evolution: where we stand and how we can move forward}",
      journal = {\aapr},
         year = 2013,
        month = feb,
       volume = {21},
          eid = {59},
        pages = {59},
          doi = {10.1007/s00159-013-0059-2},
archivePrefix = {arXiv},
       eprint = {1209.4302},
 primaryClass = {astro-ph.HE},
       adsurl = {https://ui.adsabs.harvard.edu/abs/2013A&ARv..21...59I}
}

@ARTICLE{2021A&A...648L...6P,
       author = {{Politano}, M.},
        title = "{The final orbital separation in common envelope evolution}",
      journal = {\aap},
         year = 2021,
        month = apr,
       volume = {648},
          eid = {L6},
        pages = {L6},
          doi = {10.1051/0004-6361/202140442},
archivePrefix = {arXiv},
       eprint = {2104.01487},
 primaryClass = {astro-ph.SR},
       adsurl = {https://ui.adsabs.harvard.edu/abs/2021A&A...648L...6P}
}

@ARTICLE{2026NewA..12202477N,
       author = {{Noughani}, N. and {Nordhaus}, J. and {Richmond}, M. and {Wilson}, E.~C.},
        title = "{Light curve models of convective common envelopes}",
      journal = {\na},
         year = 2026,
        month = jan,
       volume = {122},
          eid = {102477},
        pages = {102477},
          doi = {10.1016/j.newast.2025.102477},
archivePrefix = {arXiv},
       eprint = {2406.04118},
 primaryClass = {astro-ph.SR},
       adsurl = {https://ui.adsabs.harvard.edu/abs/2026NewA..12202477N}
}

@ARTICLE{2021A&A...651A.100O,
       author = {{Olejak}, A. and {Belczynski}, K. and {Ivanova}, N.},
        title = "{Impact of common envelope development criteria on the formation of LIGO/Virgo sources}",
      journal = {\aap},
         year = 2021,
        month = jul,
       volume = {651},
          eid = {A100},
        pages = {A100},
          doi = {10.1051/0004-6361/202140520},
archivePrefix = {arXiv},
       eprint = {2102.05649},
 primaryClass = {astro-ph.HE},
       adsurl = {https://ui.adsabs.harvard.edu/abs/2021A&A...651A.100O}
}

@ARTICLE{2022MNRAS.511.5462T,
       author = {{Torres}, S. and {Canals}, P. and {Jim{\'e}nez-Esteban}, F.~M. and {Rebassa-Mansergas}, A. and {Solano}, E.},
        title = "{A population synthesis fitting of the Gaia resolved white dwarf binary population within 100 pc}",
      journal = {\mnras},
         year = 2022,
        month = apr,
       volume = {511},
       number = {4},
        pages = {5462-5474},
          doi = {10.1093/mnras/stac374},
archivePrefix = {arXiv},
       eprint = {2202.04199},
 primaryClass = {astro-ph.SR},
       adsurl = {https://ui.adsabs.harvard.edu/abs/2022MNRAS.511.5462T}
}

@ARTICLE{2001A&A...368..939N,
       author = {{Nelemans}, G. and {Portegies Zwart}, S.~F. and {Verbunt}, F. and {Yungelson}, L.~R.},
        title = "{Population synthesis for double white dwarfs. II. Semi-detached systems: AM CVn stars}",
      journal = {\aap},
         year = 2001,
        month = mar,
       volume = {368},
        pages = {939-949},
          doi = {10.1051/0004-6361:20010049},
archivePrefix = {arXiv},
       eprint = {astro-ph/0101123},
 primaryClass = {astro-ph},
       adsurl = {https://ui.adsabs.harvard.edu/abs/2001A&A...368..939N}
}

@ARTICLE{2019MNRAS.488.4740P,
       author = {{Pittordis}, Charalambos and {Sutherland}, Will},
        title = "{Testing modified gravity with wide binaries in Gaia DR2}",
      journal = {\mnras},
         year = 2019,
        month = oct,
       volume = {488},
       number = {4},
        pages = {4740-4752},
          doi = {10.1093/mnras/stz1898},
archivePrefix = {arXiv},
       eprint = {1905.09619},
 primaryClass = {astro-ph.CO},
       adsurl = {https://ui.adsabs.harvard.edu/abs/2019MNRAS.488.4740P}
}

@ARTICLE{2024MNRAS.527.4573B,
       author = {{Banik}, Indranil and {Pittordis}, Charalambos and {Sutherland}, Will and {Famaey}, Benoit and {Ibata}, Rodrigo and {Mieske}, Steffen and {Zhao}, Hongsheng},
        title = "{Strong constraints on the gravitational law from Gaia DR3 wide binaries}",
      journal = {\mnras},
         year = 2024,
        month = jan,
       volume = {527},
       number = {3},
        pages = {4573-4615},
          doi = {10.1093/mnras/stad3393},
archivePrefix = {arXiv},
       eprint = {2311.03436},
 primaryClass = {astro-ph.SR},
       adsurl = {https://ui.adsabs.harvard.edu/abs/2024MNRAS.527.4573B}
}

@ARTICLE{2024MNRAS.532.2534M,
       author = {{Munday}, James and {Pelisoli}, Ingrid and {Tremblay}, P. -E. and {Marsh}, T.~R. and {Nelemans}, Gijs and {B{\'e}dard}, Antoine and {Toonen}, Silvia and {Breedt}, Elm{\'e} and {Cunningham}, Tim and {O'Brien}, Mairi W. and {Dawson}, Harry},
        title = "{The DBL Survey I: discovery of 34 double-lined double white dwarf binaries}",
      journal = {\mnras},
         year = 2024,
        month = aug,
       volume = {532},
       number = {2},
        pages = {2534-2556},
          doi = {10.1093/mnras/stae1645},
archivePrefix = {arXiv},
       eprint = {2407.02594},
 primaryClass = {astro-ph.SR},
       adsurl = {https://ui.adsabs.harvard.edu/abs/2024MNRAS.532.2534M}
}

@ARTICLE{2025MNRAS.541.3494M,
       author = {{Munday}, James and {Pelisoli}, Ingrid and {Tremblay}, Pier-Emmanuel and {Jones}, David and {Nelemans}, Gijs and {Kilic}, Mukremin and {Cunningham}, Tim and {Toonen}, Silvia and {Santos-Garc{\'\i}a}, Alejandro and {Dawson}, Harry and {Pinter}, Viktoria and {Godson}, Benjamin and {Martinez}, Llanos and {Chand}, Jaya and {Dobson}, Ross and {Jhass}, Kiran and {Shenoy}, Shravya},
        title = "{The DBL Survey II: towards a mass{\textendash}period distribution of double white dwarf binaries}",
      journal = {\mnras},
         year = 2025,
        month = aug,
       volume = {541},
       number = {4},
        pages = {3494-3512},
          doi = {10.1093/mnras/staf1198},
archivePrefix = {arXiv},
       eprint = {2507.14123},
 primaryClass = {astro-ph.SR},
       adsurl = {https://ui.adsabs.harvard.edu/abs/2025MNRAS.541.3494M}
}

@ARTICLE{2000A&A...360.1011N,
       author = {{Nelemans}, G. and {Verbunt}, F. and {Yungelson}, L.~R. and {Portegies Zwart}, Simon F.},
        title = "{Reconstructing the evolution of double helium white dwarfs: envelope loss without spiral-in}",
      journal = {\aap},
         year = 2000,
        month = aug,
       volume = {360},
        pages = {1011-1018},
          doi = {10.48550/arXiv.astro-ph/0006216},
archivePrefix = {arXiv},
       eprint = {astro-ph/0006216},
 primaryClass = {astro-ph},
       adsurl = {https://ui.adsabs.harvard.edu/abs/2000A&A...360.1011N}
}

@ARTICLE{2019ApJ...871..148L,
       author = {{Li}, Zhenwei and {Chen}, Xuefei and {Chen}, Hai-Liang and {Han}, Zhanwen},
        title = "{Formation of Extremely Low-mass White Dwarfs in Double Degenerates}",
      journal = {\apj},
         year = 2019,
        month = feb,
       volume = {871},
       number = {2},
          eid = {148},
        pages = {148},
          doi = {10.3847/1538-4357/aaf9a1},
archivePrefix = {arXiv},
       eprint = {1812.07226},
 primaryClass = {astro-ph.SR},
       adsurl = {https://ui.adsabs.harvard.edu/abs/2019ApJ...871..148L}
}

@ARTICLE{1997A&A...327..620S,
       author = {{Soberman}, G.~E. and {Phinney}, E.~S. and {van den Heuvel}, E.~P.~J.},
        title = "{Stability criteria for mass transfer in binary stellar evolution.}",
      journal = {\aap},
         year = 1997,
        month = nov,
       volume = {327},
        pages = {620-635},
          doi = {10.48550/arXiv.astro-ph/9703016},
archivePrefix = {arXiv},
       eprint = {astro-ph/9703016},
 primaryClass = {astro-ph},
       adsurl = {https://ui.adsabs.harvard.edu/abs/1997A&A...327..620S}
}

@ARTICLE{2022MNRAS.515.1228K,
       author = {{Korol}, Valeriya and {Belokurov}, Vasily and {Toonen}, Silvia},
        title = "{A gap in the double white dwarf separation distribution caused by the common-envelope evolution: astrometric evidence from Gaia}",
      journal = {\mnras},
         year = 2022,
        month = sep,
       volume = {515},
       number = {1},
        pages = {1228-1246},
          doi = {10.1093/mnras/stac1686},
archivePrefix = {arXiv},
       eprint = {2203.03659},
 primaryClass = {astro-ph.SR},
       adsurl = {https://ui.adsabs.harvard.edu/abs/2022MNRAS.515.1228K}
}

@ARTICLE{2024arXiv241117333G,
       author = {{Ge}, Hongwei and {Han}, Zhanwen},
        title = "{Mass Transfer Physics in Binary Stars and Applications in Gravitational Wave Sources}",
      journal = {arXiv e-prints},
         year = 2024,
        month = nov,
          eid = {arXiv:2411.17333},
        pages = {arXiv:2411.17333},
          doi = {10.48550/arXiv.2411.17333},
archivePrefix = {arXiv},
       eprint = {2411.17333},
 primaryClass = {astro-ph.SR},
       adsurl = {https://ui.adsabs.harvard.edu/abs/2024arXiv241117333G}
}

@ARTICLE{2016A&A...595A..35I,
       author = {{Istrate}, A.~G. and {Marchant}, P. and {Tauris}, T.~M. and {Langer}, N. and {Stancliffe}, R.~J. and {Grassitelli}, L.},
        title = "{Models of low-mass helium white dwarfs including gravitational settling, thermal and chemical diffusion, and rotational mixing}",
      journal = {\aap},
         year = 2016,
        month = oct,
       volume = {595},
          eid = {A35},
        pages = {A35},
          doi = {10.1051/0004-6361/201628874},
archivePrefix = {arXiv},
       eprint = {1606.04947},
 primaryClass = {astro-ph.SR},
       adsurl = {https://ui.adsabs.harvard.edu/abs/2016A&A...595A..35I}
}

@ARTICLE{2000MNRAS.316...84S,
       author = {{Sarna}, Marek J. and {Ergma}, Ene and {Ger{\v{s}}kevit{\v{s}}-Antipova}, Jelena},
        title = "{Cooling curves and initial models for low-mass white dwarfs (<0.25M$_{solar}$) with helium cores}",
      journal = {\mnras},
         year = 2000,
        month = jul,
       volume = {316},
       number = {1},
        pages = {84-96},
          doi = {10.1046/j.1365-8711.2000.03503.x},
archivePrefix = {arXiv},
       eprint = {astro-ph/0002261},
 primaryClass = {astro-ph},
       adsurl = {https://ui.adsabs.harvard.edu/abs/2000MNRAS.316...84S}
}

@ARTICLE{2016ApJ...818..155B,
       author = {{Brown}, Warren R. and {Gianninas}, A. and {Kilic}, Mukremin and {Kenyon}, Scott J. and {Allende Prieto}, Carlos},
        title = "{The ELM Survey. VII. Orbital Properties of Low-Mass White Dwarf Binaries}",
      journal = {\apj},
         year = 2016,
        month = feb,
       volume = {818},
       number = {2},
          eid = {155},
        pages = {155},
          doi = {10.3847/0004-637X/818/2/155},
archivePrefix = {arXiv},
       eprint = {1604.04268},
 primaryClass = {astro-ph.SR},
       adsurl = {https://ui.adsabs.harvard.edu/abs/2016ApJ...818..155B}
}

@ARTICLE{2010ApJ...717.1006R,
       author = {{Ruiter}, Ashley J. and {Belczynski}, Krzysztof and {Benacquista}, Matthew and {Larson}, Shane L. and {Williams}, Gabriel},
        title = "{The LISA Gravitational Wave Foreground: A Study of Double White Dwarfs}",
      journal = {\apj},
         year = 2010,
        month = jul,
       volume = {717},
       number = {2},
        pages = {1006-1021},
          doi = {10.1088/0004-637X/717/2/1006},
archivePrefix = {arXiv},
       eprint = {0705.3272},
 primaryClass = {astro-ph},
       adsurl = {https://ui.adsabs.harvard.edu/abs/2010ApJ...717.1006R}
}

@ARTICLE{2023MNRAS.524..245R,
       author = {{Romero-Shaw}, Isobel and {Hirai}, Ryosuke and {Bahramian}, Arash and {Willcox}, Reinhold and {Mandel}, Ilya},
        title = "{Rapid population synthesis of black hole high-mass X-ray binaries: implications for binary stellar evolution}",
      journal = {\mnras},
         year = 2023,
        month = sep,
       volume = {524},
       number = {1},
        pages = {245-259},
          doi = {10.1093/mnras/stad1732},
archivePrefix = {arXiv},
       eprint = {2303.05375},
 primaryClass = {astro-ph.HE},
       adsurl = {https://ui.adsabs.harvard.edu/abs/2023MNRAS.524..245R}
}

@ARTICLE{2024MNRAS.534.3506R,
       author = {{Rauf}, Liana and {Howlett}, Cullan and {Stevenson}, Simon and {Riley}, Jeff and {Willcox}, Reinhold},
        title = "{A trifecta of modelling tools: a Bayesian binary black hole model selection combining population synthesis and galaxy formation models}",
      journal = {\mnras},
         year = 2024,
        month = nov,
       volume = {534},
       number = {4},
        pages = {3506-3539},
          doi = {10.1093/mnras/stae2288},
archivePrefix = {arXiv},
       eprint = {2406.11885},
 primaryClass = {astro-ph.HE},
       adsurl = {https://ui.adsabs.harvard.edu/abs/2024MNRAS.534.3506R}
}

@ARTICLE{2024A&A...682A..33B,
       author = {{Belloni}, Diogo and {Schreiber}, Matthias R. and {Moe}, Maxwell and {El-Badry}, Kareem and {Shen}, Ken J.},
        title = "{Evidence for saturated and disrupted magnetic braking from samples of detached close binaries with M and K dwarfs}",
      journal = {\aap},
         year = 2024,
        month = feb,
       volume = {682},
          eid = {A33},
        pages = {A33},
          doi = {10.1051/0004-6361/202347931},
archivePrefix = {arXiv},
       eprint = {2311.04309},
 primaryClass = {astro-ph.SR},
       adsurl = {https://ui.adsabs.harvard.edu/abs/2024A&A...682A..33B}
}

@ARTICLE{2001A&A...365..491N,
       author = {{Nelemans}, G. and {Yungelson}, L.~R. and {Portegies Zwart}, S.~F. and {Verbunt}, F.},
        title = "{Population synthesis for double white dwarfs . I. Close detached systems}",
      journal = {\aap},
         year = 2001,
        month = jan,
       volume = {365},
        pages = {491-507},
          doi = {10.1051/0004-6361:20000147},
archivePrefix = {arXiv},
       eprint = {astro-ph/0010457},
 primaryClass = {astro-ph},
       adsurl = {https://ui.adsabs.harvard.edu/abs/2001A&A...365..491N}
}

@ARTICLE{2010A&A...521A..85Y,
       author = {{Yu}, S. and {Jeffery}, C.~S.},
        title = "{The gravitational wave signal from diverse populations of double white dwarf binaries in the Galaxy}",
      journal = {\aap},
         year = 2010,
        month = oct,
       volume = {521},
          eid = {A85},
        pages = {A85},
          doi = {10.1051/0004-6361/201014827},
archivePrefix = {arXiv},
       eprint = {1007.4267},
 primaryClass = {astro-ph.SR},
       adsurl = {https://ui.adsabs.harvard.edu/abs/2010A&A...521A..85Y}
}

@ARTICLE{2023A&A...672A..54S,
       author = {{Siegert}, Thomas and {Pleintinger}, Moritz M.~M. and {Diehl}, Roland and {Krause}, Martin G.~H. and {Greiner}, Jochen and {Weinberger}, Christoph},
        title = "{Galactic population synthesis of radioactive nucleosynthesis ejecta}",
      journal = {\aap},
         year = 2023,
        month = apr,
       volume = {672},
          eid = {A54},
        pages = {A54},
          doi = {10.1051/0004-6361/202244457},
archivePrefix = {arXiv},
       eprint = {2301.10192},
 primaryClass = {astro-ph.GA},
       adsurl = {https://ui.adsabs.harvard.edu/abs/2023A&A...672A..54S}
}

@ARTICLE{1988ApJ...334..688L,
       author = {{Lee}, Hyung Mok and {Nelson}, Lorne A.},
        title = "{Stellar Encounters in the Galactic Center: Formation of Massive Stars and Close Binaries}",
      journal = {\apj},
         year = 1988,
        month = nov,
       volume = {334},
        pages = {688},
          doi = {10.1086/166870},
       adsurl = {https://ui.adsabs.harvard.edu/abs/1988ApJ...334..688L}
}

\end{document}